\documentclass[aps,pra,reprint,showkeys,longbibliography]{revtex4-2} 
\usepackage{amsmath}
\usepackage{amssymb}
\usepackage{graphicx} 
\usepackage{bm} 
\usepackage{xcolor}
\usepackage[toc,page]{appendix}
\usepackage{braket}
\usepackage{mathrsfs}

\makeatletter

\@ifclasswith{revtex4-2}{draft}{%
    \usepackage[notref,notcite]{showkeys}
    
}{}

\makeatother

\newcommand{\nn}{\nonumber}

\begin{document}

\title{Quantum observables for probabilistic classical particles}

\author{Christof Wetterich}
\affiliation{Institut f\"ur Theoretische Physik\\
    Universit\"at Heidelberg\\
    Philosophenweg 16, D-69120 Heidelberg}

\begin{abstract}
The classical observables of position and momentum are not well adapted to particles in a microphysical situation where typical probability distributions are characterized by a substantial dispersion.
We propose the use of more robust quantum observables for probabilistic classical particles.
The quantum observables are statistical observables which do not take fixed values for a given classical position and momentum. 
Solutions of the Liouville equation are discussed in the quantum formalism for classical statistics.
Statistical observables are represented by non-commuting operators.
No classical correlation function is defined for these observables and Bell's inequalities do not apply.
We demonstrate for a general potential how a quantum system emerges from classical statistics.
For the particular cases of a harmonic potential and a Coulomb potential we investigate subsystems which describe all features of a quantum particle.
This covers the discrete energy spectrum of the hydrogen atom and quantum harmonic oscillator.
We discuss the interference for the double-slit experiment.
Conserved statistical observables may also be relevant for the probabilistic dynamics of dust or planets.
\end{abstract}

\maketitle

\section{Introduction}

\indent The classical observables of position and momentum for a particle are an idealisation which assumes that both can be measured precisely.
If one would like to investigate the behavior a classical particle in a microphysical situation, as for electrons in the Coulomb potential of an atom, a probabilistic description by the Liouville equation would be appropriate.
The probability distribution for the particle would typically be characterized by a substantial dispersion in momentum and position.
While the classical position and momentum observables remain well defined, they are not very appropriate in practice.
Repeated measurements of position would (almost) never yield a sharp value.
Probability distributions with a sharp value of position are mathematically possible, but not realized in practice.
Due to the spread in the momentum distribution such sharp probability distributions broaden on a time scale much shorter than the one for the measurement.

\indent A more robust definition of a position observable takes into account the roughness of the probability distribution in momentum.
Such an observable has an intrinsic uncertainty given by the variation of the momentum distribution.
It is a statistical observable since properties of the probability distribution matter.
Its status is somewhat analogous to temperature or pressure in classical statistical equilibrium.
Temperature and pressure are observables which can well be measured, but they have no fixed value for a given point in phase space for the particles.
For the probabilistic description of a single classical pointlike particle more robust statistical observables could be realized by the quantum observables \cite{CWQP1,CWQP2} for position and momentum.
They involve properties of the probability distribution and do not have fixed values for the points in phase space.

\indent We investigate solutions of the Liouville equation in the quantum formalism for classical statistics \cite{CWQFC}.
The quantum observables for position and momentum, as well as the quantum energy, play a crucial role for the dynamics.
The Liouville equation can be interpreted as the Schrödinger equation for a particular type of quantum system. 
This demonstrates how quantum systems are embedded in classical statistics.
We do not want to suggest that real electrons in a Coulomb potential are described by probabilistic classical particles.
Real electrons obey the Schrödinger equation for a quantum system which is simpler than the one for the classical point particle.
Our aim is rather a demonstration that all conceptual properties of quantum systems are already realized for the classical probabilistic description of point particles.
In the particular case of a harmonic potential or for free particles the quantum system for the classical point particle coincides with the standard quantum particle once one uses the quantum observables.

\indent The probabilistic description of classical particles by the Liouville equation is not confined to microphysics.
It matters as well for macroscopic physics as the distribution of dust or planets around stars.
Since we use the same quantum formalism both for microphysics and macrophysics a comparison becomes straightforward.
Among the statistical observables we identify conserved quantities.
They may be useful for understanding the dynamics of probabilistic systems on all scales.

\indent The discussion of quantum features for the classical Liouville equation has been pioneered by Koopman \cite{KOOP} and von Neumann \cite{VNEU}, with many interesting subsequent applications of an operator formalism \cite{MAU, GOMA, BON, PIAS, GOMA2, KLE, MMS, AME, KPM, GIA, CHRU, MEZI, DW, GBT}, see also \cite{GRT, KOBR, KRBG, GAL, BICE, BER, KAN, MAMA, NIC, VOLO, ACKA, GOZ1, CWET1, CWET2, NAKL} for some related developments.
This approach was not considered, however, as a realisation of a true quantum system since all observables were commuting, not reproducing the crucial non-commuting features of quantum mechanics.
The quantum observables proposed in the present work overcome this shortcoming.
The operators associated to the quantum position and momentum obey the standard commutation relation of quantum mechanics.
A second important difference between the quantum formalism for classical statistics and the approach of Koopman and von Neumann is the use of a real classical wave function \cite{CWQMCS1, CWQP1, CWQFC, CWIT} whose components are given by the square roots of the probabilities.
A complex Schrödinger equation can be obtained by a Fourier transform.
The real character of the classical wave function is reflected in this setting by an important selection rule which is crucial for the understanding of the dynamics.  
In contrast to the approach of Koopman and von Neumann the phases of this complex wave function matter for the dynamics and expectation values of observables, reflecting the characteristic interference effect of quantum mechanics.

\indent One possible description of the probabilistic classical particle employs a combined quantum system for a quantum particle and its mirror particle.
Particle and mirror particle obtain from each other by time reversal.
The eigenvalues of the Hamiltonian come in pairs with positive and negative energies, related by time reversal.
This quantum system is stable, nevertheless.
The selection rule induces strong relations between the quantum particle and its mirror particle, which guarantee the stability expected for the probability distribution of a classical particle in a potential.

\indent For the particular case of a harmonic potential or a free particle the mirror particle decouples from the quantum particle.
In this case the mirror particle plays no role for the dynamics and observations in the subsystem for the quantum particle.
For generic potentials the quantum system for the probabilistic classical particle needs to take into account the interaction between the quantum particle and its mirror particle.
For small interactions the picture of a quantum particle and its mirror particle remains useful.

\indent For large deviations from a harmonic potential different quantum subsystems are better suited for an understanding of the dynamics.
For the Coulomb potential we propose a setting of statistical observables which reflects the $SO(4)$-symmetry.
The quantum observables obey the corresponding non-trivial commutation relations.
We find a subsystem that reproduces all properties of the quantum mechanics for a spinless electron in the hydrogen atom.

\indent In sect.~\ref{sec:II} we describe the quantum formalism for a probabilistic classical particle in an arbitrary potential.
Sect.~\ref{sec:III} provides a first discussion of the central concept of statistical observables, with focus on quantum position and momentum in sec.~\ref{sec:IV}.
Sect.~\ref{sec:V} discusses the classical statistical setting which realizes the free quantum particle with the characteristic dispersive propagation, and the quantum harmonic oscillator with its discrete equidistant frequency spectrum.
We also address the boundary problem for the free particle and discuss the double-slit experiment for the probabilistic classical particle.
In sect.~\ref{sec:VI} we introduce the quantum angular momentum.
An example for eigenstates of the quantum angular momentum is presented in detail in sect.~\ref{sec:VII}.
In sect.~\ref{sec:VIII} we discuss general quantum subsystems based on a closed algebra of statistical or classical observables.
This provides a bridge to the historical quantisation procedures based on Poisson brackets.
In sect.~\ref{sec:IX} we turn to the probabilistic classical particle in the Coulomb or Newton potential.
Sect.~\ref{sec:X} shows how the quantum hydrogen atom obtains from this setting as a subsystem.
In sect.~\ref{sec:XI} we address the issue of periodic probability distributions for the classical particle.
Sect.~\ref{sec:XII} contains our conclusions.

\section{Quantum system for probabilistic classical particle}
\label{sec:II}
\subsection*{Variables and scales}

\indent Our starting point is the Liouville equation
\begin{align}
\partial_t w &= - \mathcal{L} w\,, \nn \\
\mathcal{L} &= v \partial_z + F(z)\partial_v\,,
\label{eq:1}
\end{align}
which governs the time evolution of the probability distribution
$w(t;z,v)$ for a particle with position vector $z$ and momentum vector $v$.
Eq.~\eqref{eq:1} involves scalar products, 
$v \partial_z = \sum_k v_k \frac{\partial}{\partial z_k}$, $F(z)\partial_v = \sum_k F_k(z)\frac{\partial}{\partial v_k}$.
The number of components is arbitrary. In the limit of infinite-dimensional
vectors eq.~\eqref{eq:1} can describe fields, $z_k \to \phi(\vec x)$, $v_k = \pi(\vec x)$ \cite{CW26}.
We focus here on a single particle in a potential, where the dimension of
the vectors corresponds to the dimension of space.

\indent We use non-relativistic units with $\hbar =1$ and with particle mass $m=1$, for which momentum and velocity coincide.
For $\hbar=1$ momentum is measured in inverse length units, and energy in inverse time units.
In these units the mass has dimension time over length squared.
Setting $m=1$ measures time in units of length squared, or length in units of square root of time.
We employ time, or more precisely $mt$, to set the remaining unit.
In these conventions one has the units $z \sim t^{1/2}$, $v \sim t^{-1/2}$, $F \sim t^{-3/2}$, $\mathcal{L} \sim t^{-1}$.
Reinserting a more standard unit for $m$ the physical quantities obtain from our variables by
\begin{align}
    \label{eq:1A}
    t_{\text{ph}} = \frac{t}{m}\,, \quad x_{\text{ph}} = \frac{z}{m}\,, \quad p =mv \,, \quad F_{\text{ph}} = m^2 F\,.
\end{align}

\indent Inserting these relations, the Liouville equation \eqref{eq:1} takes the standard form for a particle with mass $m$,
\begin{equation}
    \label{eq:1B}
    \frac{\partial}{ \partial (mt_{\text{ph}})} w =  -\left( \frac{p}{m} \frac{\partial}{\partial(mx_{\text{ph}})}  + \frac{F_{\text{ph}}}{m^2} \frac{\partial}{\partial(p/m)} \right)w\,.
\end{equation}
In the limit of the sharp probability distribution for a point particle the Liouville equation yields
Newton's equation for the position $X$ or $Z=mX$ of the point particle
\begin{equation}
    \label{eq:1C}
    \frac{\partial^2}{\partial(mt_{\text{ph}})^2} (mX) = \frac{F_{\text{ph}}}{m^2} \,\, \hat{=} \,\, \frac{\partial^2 Z}{\partial t^2} = F \,.
\end{equation}
The energy and potential are rescaled by
\begin{equation}
    \label{eq:1D}
    E_{\text{ph}} = \frac{E}{m} \,, \quad V_{\text{ph}} = \frac{V}{m} \,.
\end{equation}
In terms of units of length and physical time the unit of $t=m t_{\text{ph}}$ is $v^{-2} \sim t^2_{\text{ph}}/l^2$,
with units for $m \sim t_{\text{ph}}/l^2$.

\indent Our units allow for a unified description independent of the mass of particles, which may range from planets to electrons.
For certain shapes of the potential, as the Coulomb potential, we will observe an additional scale symmetry.
This will further help a universal description over all scales.

\subsection*{Wave function for classical statistics}

\indent Instead of working with $w$ we employ the classical wave function $q$ \cite{CWQP1, CWIT}
defined as its square root up to an arbitrary sign, and obeying
\begin{align}
w(t) &= q^2(t)\,, \quad
\partial_t q = - \mathcal{L} q \,.
\label{eq:2}
\end{align}
The wave function is a real unit vector,
\begin{equation}
\int\! dz \int\! dv \, q^2(t;z,v) = 1 \,.
\label{eq:3}
\end{equation}
This guarantees a positive and normalized probability distribution $w$.
(Momentum type integrations as $\int_v$ include factors of $(1/2\pi)$ as frequently used for Fourier transforms.)

\indent An important advantage of the use of wave functions is the possibility
to perform basis transformations. 
We consider the Fourier transform
\begin{equation}
\varphi(t;z,s) = \int_v \, e^{isv}\, q(t;z,v) \,.
\label{eq:4}
\end{equation}
This wave function is complex. 
Since $q$ is real it obeys an important selection rule
\begin{equation}
\varphi^\ast(t;z,s) = \varphi(t;z,-s) \,.
\label{eq:5}
\end{equation}
The Liouville equation \eqref{eq:1} translates to the complex Schr\"odinger
equation with hermitian Hamiltonian $H$,
\begin{align}
i\partial_t \varphi &= H\varphi, \quad H^\dagger = H\,, \nn \\
H &= - \partial_z \partial_s - s F(z) \,.
\label{eq:6}
\end{align}
The eigenvalues of $H$ come in pairs with positive and negative energies.
They are mapped to each other by time reversal $T$ which changes the sign
of $v$ and $s$. 
In contrast to the formulation of Koopman and von Neumann the phases of $\varphi$ do matter for the dynamics.

\subsection*{Quantum particle and mirror particle}

\indent The quantum system defined by $H$ can be viewed as describing a quantum
particle at position $x$ coupled to a mirror particle at position $y$, where
\begin{equation}
z = \frac{x+y}{2}\,, \quad s = x-y \,.
\label{eq:8}
\end{equation}
The selection rule relates the particle and the mirror particle by
\begin{equation}
\varphi^\ast(t;x,y) = \varphi(t;y,x) \,.
\label{eq:10}
\end{equation}
In terms of $x$ and $y$, the Hamiltonian \eqref{eq:6} takes the form
\begin{equation}
H = -\frac{1}{2}\left(\partial_x^2 - \partial_y^2\right)
    - (x-y)\,F\!\left(\frac{x+y}{2}\right) \,.
\label{eq:9}
\end{equation}
We split it into its parts $H^{(p)}$ for the particle, $H^{(m)}$ for the mirror particle, and an interaction $H^{(\mathrm{int})}$,
\begin{align}
H &= H^{(p)} - H^{(m)} + H^{(\mathrm{int})}\,, \nn \\
H^{(p)} &= -\frac{1}{2}\partial_x^2 + \tilde V(x)\,, \quad
H^{(m)} = -\frac{1}{2}\partial_y^2 + \tilde V(y)\,.
\label{eq:11}
\end{align}
Here the potential $\tilde V$ is defined by
\begin{equation}
\tilde V(x) = -xF\left(\frac{x}{2}\right)\,, \quad \tilde V(y) = -yF\left(\frac{y}{2}\right)\,,
\label{eq:12}
\end{equation}
and the interaction part reads
\begin{equation}
H^{(\mathrm{int})}
= xF\left(\frac{x}{2}\right) - yF\left(\frac{y}{2}\right) - (x-y)\,F\!\left(\frac{x+y}{2}\right) \,.
\label{eq:13}
\end{equation}

\indent For a linear force or harmonic potential the interaction part vanishes.
For an anharmonic potential in one dimension, with
\begin{equation}
F(z) = -\left(cz + dz^3\right) \,,
\label{eq:14}
\end{equation}
one has
\begin{align}
\tilde V(x) &= \frac{c}{2}x^2 + \frac{d}{8}x^4\,, \quad
\tilde V(y) = \frac{c}{2}y^2 + \frac{d}{8}y^4\,, \nn \\
H_{\mathrm{int}} &= \frac{d}{4}xy\,(x^2-y^2) \,.
\label{eq:15}
\end{align}
In general, the generalized potential $\tilde V$ does not equal the
classical potential $V_{\mathrm{cl}}$ which can be defined if $F$ is
a gradient,
\begin{equation}
F(z) = - \partial_z V_{\mathrm{cl}}(z) \,.
\label{eq:16}
\end{equation}
For the anharmonic force \eqref{eq:14} the coefficient of the quartic term in
$V_{\mathrm{cl}} = cz^2/2 + dz^4/4$ differs from the one in the potential $\tilde V$.

\indent The Hamiltonian for the mirror particle is the negative of the Hamiltonian
for the particle. 
This reflects that all eigenvalues of $H$ come in pairs with opposite sign. 
At first sight one may have the suspicion that the quantum system is unstable in view of its Hamiltonian not being bounded from below and above. 
The mirror particle is, however, not independent of the particle, being related to it by time reversal, as expressed by the selection rule \eqref{eq:10}. 
This is somewhat reminiscent of the positive and negative eigenvalues of the Dirac equation. 
In contrast to the antiparticles in quantum field theory, the mirror particle is not an independently propagating degree of freedom. 
Of course, we could infer the stability of the quantum system from the stability of the equivalent Liouville system.
Understanding the stability of the quantum system with unbounded Hamiltonian will reveal important relations between the particle and the mirror particle.

\subsection*{Quantum rule for expectation values}

\indent As familiar in quantum mechanics one associates an operator $\hat A$ to an observable $A$ and computes the expectation value by the quantum rule
\begin{align}
\langle A(t)\rangle
&= \varphi^\dagger(t)\,\hat A\,\varphi(t) \nn \\
&= \int\! dx \int\! dy \,
\varphi^\dagger(t;x,y)\,\hat A\,\varphi(t;x,y) \,.
\label{eq:17}
\end{align}
(For simplicity we focus on observables with time-independent $\hat A$.)
The observables include the classical observables $z$ and $v$ which are expressed by the commuting operators
\begin{equation}
\hat z = \frac{x+y}{2}\,, \quad
\hat v = -i\partial_s = -\frac{i}{2}\left(\partial_x-\partial_y\right)\,.
\label{eq:18}
\end{equation}
Classical observables are arbitrary functions $B(z,v)$, expressed by operators $\hat A = B(\hat z,\hat v)$. 
They all commute among themselves.
The equivalence of the quantum rule \eqref{eq:17} with the classical probabilistic definition,
\begin{equation}
\langle B(z,v)\rangle
= \int_{z,v} \, B(z,v)\, w(t;z,v) \,,
\label{eq:19}
\end{equation}
is easily established by the inverse of the Fourier transform \eqref{eq:4} and the relation \eqref{eq:2}.
An important classical observable is the classical energy
\begin{equation}
E_{\mathrm{cl}} = \frac{v^2}{2} + V_{\mathrm{cl}}(z) \,.
\label{eq:20}
\end{equation}
The associated operator $\hat E_{\mathrm{cl}}$ differs from the quantum Hamiltonian $H$. 
Static solutions of the Liouville equation \eqref{eq:2} obtain for arbitrary classical wave functions depending only on $E_{\mathrm{cl}}(z,v)$,
\begin{equation}
\partial_t\, q\!\left(E_{\mathrm{cl}}(z,v)\right) = 0 \,.
\label{eq:21}
\end{equation}

\section{Statistical observables}
\label{sec:III}

\indent Besides the classical observables, an important role is played by statistical observables. 
They characterize properties of the probability distribution $w$ or the associated wave function $q$ or $\varphi$.
Similar to temperature in classical statistical equilibrium systems, they do not take a fixed value for a given point in phase space $(z,v)$.
The properties measured by the expectation values of statistical observables, as periodicity or roughness of the probability distribution, characterize only the probabilistic information about the system at a given time.
Classical correlation functions are not defined for pairs of statistical
observables $A$ and $B$. 
There are no simultaneous probabilities for $A$ taking the value $a$ and $B$ taking the value $b$. 
Bell's inequalities \cite{BEL1, CHSH} for classical correlation functions therefore find no application to pairs of statistical observables. 
One may still be able to define correlations for such pairs \cite{CWQMCS1, CWPW, CWPW2}, but they are not given by classical correlation functions.
We will see that statistical observables are represented by operators which do not commute among themselves and with the operators for all classical observables. 
While in macrophysics only a few statistical observables such as temperature, pressure or the chemical potential are used in practice, the statistical observables play a crucial role for the understanding of microphysics.

\subsection*{Quantum Hamiltonian}

\indent An important statistical observable is the quantum energy, represented by the Hamiltonian $H$ of the quantum system. 
It does not commute with $\hat z$ or $\hat v$,
\begin{equation}
[H,\hat z] = - i \hat v\,, \quad
[H,\hat v] = - i F(\hat z)\,,
\label{eq:22}
\end{equation}
and can therefore not have a fixed value for a given phase-space point $(z,v)$.
The Hamiltonian dictates the time evolution of expectation values according to
\begin{equation}
\partial_t \langle A(t)\rangle = i\,\varphi^\dagger [H,\hat A]\varphi \,.
\label{eq:23}
\end{equation}
If an operator $\hat A$ commutes with $H$ the associated observable is a
conserved quantity in the sense that its expectation value is time-independent. 
The classical energy is a conserved quantity,
\begin{equation}
[H,\hat E_{\mathrm{cl}}] = 0 \,.
\label{eq:24}
\end{equation}
This is most easily seen in the basis with the real wave function $q$,
where $H$ is directly related to the Liouville operator, $H = - i \mathcal{L}$.
The quantum energy is conserved as well, since $H$ obviously commutes with itself.
Conserved quantities are important tools for the understanding of the
dynamics of a system.
Arbitrary functions of a conserved quantity are conserved. 
This holds both for classical and statistical observables. 

\indent The expectation value of $H$ vanishes
\begin{equation}
\langle H \rangle = 0 \,.
\label{eq:25}
\end{equation}
This follows most directly from partial integration of the relation
\begin{align}
\langle H \rangle &= q^T H\,q= - i\, q^T \mathcal{L}\, q \nn \\
&= i \int_{z,v} q(z,v)\,
\bigl(v\partial_z + F(z)\partial_v\bigr)\, q(z,v) \,.
\label{eq:26}
\end{align}
The only eigenfunctions of $H$ which are compatible with the selection rule
\eqref{eq:5} are the ones for vanishing quantum energy. 
For any eigenfunction $\varphi_n$,
\begin{equation}
H \varphi_n(z,s) = E_n \varphi_n(z,s) \,,
\label{eq:26A}
\end{equation}
there exists an eigenfunction with opposite $E_n$,
\begin{equation}
H \varphi_n(z,-s) = - E_n \varphi_n(z,-s) \,,
\label{eq:26B}
\end{equation}
which obtains by time reversal $s \to -s$.
In the basis \eqref{eq:6} $H$ is a real operator, $H^\ast = H$. 
Then the complex conjugate of eq.~\eqref{eq:26A} reads
\begin{equation}
H \varphi_n^\ast(z,s) = E_n \varphi_n^\ast(z,s)\,, 
\label{eq:26C}
\end{equation}
and the selection rule \eqref{eq:5} entails
\begin{equation}
H \varphi_n(z,-s) = E_n \varphi_n(z,-s)\,. 
\label{eq:26D}
\end{equation}
Compatibility with eq.~\eqref{eq:26B} is given only for $E_n = 0$.

\indent In view of this property and eq.~\eqref{eq:25} one may think that $H$ is not a particularly interesting conserved quantity. 
Strictly speaking $H$ does not act within the Hilbert space for the real wave function $q$ or the complex wave function $\varphi$.
It may therefore not be considered as a genuine statistical observable.
However, arbitrary even functions of $H$ are conserved quantities which can have a non-zero expectation value.
They are statistical observables.
As an example we may take the squared quantum energy for a particle in one space-dimension,
\begin{align}
H^2 &= \partial_z^2 \partial_s^2
+ \bigl[2F(z)\partial_z + \partial_z F(z)\bigr]\, s\partial_s \nn \\
&\quad + s^2 F^2(z) + \partial_z F(z) + F(z)\partial_z \,.
\label{eq:28}
\end{align}
In general, the energy dispersion,
\begin{equation}
\langle H^2\rangle = \bigl\langle (H-\langle H\rangle)^2 \bigr\rangle \ge 0 \,,
\label{eq:29}
\end{equation}
is a positive quantity. 
It vanishes only for an eigenstate of $H$. 
The conservation of $\langle H^2\rangle$ is an important ingredient for the stability of the quantum system.
With unbounded $H$, the conservation of $H$ is not sufficient to forbid instabilities where energy eigenstates with large positive
and negative energies are more and more populated with increasing time. 
The conservation of $\langle H^2\rangle$ forbids this type of instability.
A second important use of $H$ is the construction of probability distributions with a periodic time evolution for arbitrary potentials.
We will turn to this topic later.

\subsection*{Roughness observables}

\indent Eq.~\eqref{eq:28} involves the statistical observables $s$ and $r$.  
One possible choice represents them by the operators
\begin{equation}
\hat s = s\,, \quad \hat r = - i \partial_z \,.
\label{eq:29A}
\end{equation}
The selection rule \eqref{eq:5} imposes important restrictions on the expectation
values of these statistical observables. 
A first example is the relation
\begin{align}
\langle G(z,s)\rangle
&= \varphi^\dagger G(\hat z,\hat s)\varphi \nn\\
&= \int_{z,s} \varphi^\ast(z,s)\,G(z,s)\,\varphi(z,s) \nn\\
&= \int_{z,s} \varphi(z,-s)\,G(z,s)\,\varphi^\ast(z,-s) \nn\\
&= \langle G(z,-s)\rangle\,.
\label{eq:29B}
\end{align}
For all functions $G(z,s)$ which are odd under a reflection
$s \leftrightarrow -s$ or $x \leftrightarrow y$, the expectation value
vanishes, in particular
\begin{equation}
\langle s\rangle = 0 \,.
\label{eq:29C}
\end{equation}
Similarly, one finds by a variable change $s \to -s$ and partial
integration for $z$,
\begin{align}
\langle r^n J(s)\rangle
&=\int_{z,s} \varphi(z,-s)\,(-i\partial_z)^n\,J(s)\,\varphi(z,s) \nn \\
&=(-1)^n \langle r^n J(-s)\rangle \,.
\label{eq:29D}
\end{align}
or similarly
\begin{equation}
\langle rv\rangle = - \langle \partial_z \partial_s \rangle = 0 \,.
\label{eq:29E}
\end{equation}
One may use these relations to verify $\langle H\rangle = 0$ in the complex
picture \eqref{eq:6}.

\indent Similar to $H$ the roughness observables are not genuine observables.
Their squares, and other combinations are, however, important statistical observables.
The statistical observable $s^2$ measures the roughness of the momentum
distribution of the classical particle,
\begin{align}
\hat s &= i\partial_v\,, \nn \\
\langle s^2\rangle
&= - \int_{z,v} q(z,v)\,\partial_v^2 q(z,v) \nn \\
&= \int_{z,v} \bigl(\partial_v q(z,v)\bigr)^2 \,.
\label{eq:43}
\end{align}
Thus $\langle s^2\rangle$ vanishes only if the probability distribution
$w$, and therefore also the classical wave function $q$, does not depend on
the momentum $v$. 
Otherwise $\langle s^2\rangle$ is strictly positive.
Similarly, $r^2$ measures the roughness of the position distribution.
The roughness observables play an important role for the quantum position and momentum.

\section{Quantum position and momentum}
\label{sec:IV}

\indent Important statistical observables are the position and momentum of the
quantum particle.
One possible choice represents them by the operators
\begin{equation}
\hat x = x\,, \quad \hat p = - i \partial_x \,.
\label{eq:30}
\end{equation}
This pair of observables does not commute
\begin{equation}
[\hat x_k,\hat p_l] = i \delta_{kl} \,.
\label{eq:31}
\end{equation}
The evolution of $\langle x\rangle$ and $\langle p\rangle$ are given by
\begin{align}
\partial_t \langle x\rangle &= i\langle [H,\hat x]\rangle = \langle p\rangle\,,
\nn \\
\partial_t \langle p\rangle
&= i\langle [H,\hat p]\rangle
= - \bigl\langle \partial_x \bigl(\tilde V(x)+H_{\mathrm{int}}(x,y)\bigr)\bigr\rangle \,.
\label{eq:32}
\end{align}
For the mirror particle we have similar relations
\begin{equation}
\hat y = y, \quad \hat q = - i \partial_y, \quad
[\hat y_k,\hat q_l] = i \delta_{kl} \,,
\label{eq:33}
\end{equation}
and
\begin{align}
\partial_t \langle y\rangle &= - \langle q\rangle\,, \nn\\
\partial_t \langle q\rangle
&= \bigl\langle \partial_y \bigl(\tilde V(y)-H_{\mathrm{int}}(x,y)\bigr)\bigr\rangle \,.
\label{eq:34}
\end{align}
Using the relations \eqref{eq:29B} \eqref{eq:29D} one has
\begin{align}
\langle x\rangle - \langle y\rangle &= \langle s\rangle = 0\,, \nn\\
\langle p\rangle - \langle q\rangle
&= \langle -2i\partial_s\rangle = 2\langle v\rangle\,, \nn\\
\langle p\rangle + \langle q\rangle
&= \langle r \rangle = \langle -i\partial_z\rangle = 0\,.
\label{eq:35}
\end{align}
The expectation values for the positions of the particle and the mirror particle coincide for all $t$
\begin{equation}
\langle x(t)\rangle = \langle y(t)\rangle\,, \quad
\partial_t \langle x\rangle = \partial_t \langle y\rangle = \langle v\rangle \,.
\label{eq:36}
\end{equation}
With
\begin{equation}
\langle p\rangle = - \langle q\rangle = \langle v\rangle \,,
\label{eq:37}
\end{equation}
the expectation value of the position of the quantum particle and the
mirror particle coincide for all $t$ with the expectation value $\langle z\rangle$
of the classical particle.

\indent On the level of fluctuations around the expectation value the quantum
particle differs from the classical particle. 
The selection rule implies
\begin{equation}
\langle x^2\rangle - \langle y^2\rangle = \langle 2zs\rangle = 0 \,,
\label{eq:38}
\end{equation}
or
\begin{equation}
\langle x^2\rangle = \langle y^2\rangle
= \langle z^2\rangle + \frac{1}{4}\langle s^2\rangle \,.
\label{eq:39}
\end{equation}
The fluctuations of $x$ are larger than the fluctuations of $z$ whenever
$\langle s^2\rangle > 0$,
\begin{equation}
\bigl\langle (x-\langle x\rangle)^2 \bigr\rangle
- \bigl\langle (z-\langle z\rangle)^2 \bigr\rangle
= \frac{1}{4}\langle s^2\rangle \,.
\label{eq:40}
\end{equation}
One obtains the evolution equations
\begin{align}
\partial_t \langle x^2\rangle
&= \bigl\langle i[H,\hat x^2]\bigr\rangle
= \langle \hat x\hat p + \hat p \hat x\rangle\,, \nn \\
\partial_t \langle y^2\rangle
&= \bigl\langle i[H,\hat y^2]\bigr\rangle
= - \langle \hat y\hat q + \hat q \hat y\rangle \,,
\label{eq:41}
\end{align}
where one can identify the right hand sides.
For $\langle s^2(t) \rangle$ the evolution obeys
\begin{equation}
\partial_t \langle s^2\rangle = 2\langle rs\rangle .
\label{eq:42}
\end{equation}

\indent For possible measurements in microphysical systems the quantum position and momentum are more robust
observables than the classical position and momentum.
The choice of \eqref{eq:30} for the quantum position and momentum is not unique.
There are other possible choices leading to the quantum relation \eqref{eq:31}.
We will encounter a different choice for the quantum particle in the Coulomb potential.

\section{Free quantum particle and quantum particle in a harmonic potential}
\label{sec:V}

\indent Consider a linear force,
\begin{equation}
F_k(z) = -\bigl(c_{k\ell}z_\ell + b_k\bigr)\,,
\label{eq:FQ1}
\end{equation}
which corresponds to a quadratic potential,
\begin{equation}
V(z)=V_{\mathrm{cl}}(z)=\tilde V(z)
=
a+b_k z_k+\frac{1}{2}c_{k\ell}z_k z_\ell\,.
\label{eq:FQ2}
\end{equation}
In this case the interaction term $H^{(\mathrm{int})}$ in eq.~\eqref{eq:11} vanishes. 
We focus first on a direct product wave function obeying the constraint \eqref{eq:10},
\begin{equation}
\varphi(x,y)=\psi(x)\psi^*(y)\,.
\label{eq:FQ3}
\end{equation}
The wave function of the subsystem $\psi(x)$ obeys the Schr\"odinger equation
\begin{equation}
i\partial_t\psi
=
H^{(p)}\psi
=
\left(
-\frac{1}{2}\partial_x^2+V(x)
\right)\psi\,.
\label{eq:FQ3A}
\end{equation}

\indent For this setting the mirror particle decouples from the quantum particle. 
As long as we consider only observables built from the quantum position and momentum \eqref{eq:30} the mirror particle plays no role. 
The dynamics for the quantum particle obeys the standard Schr\"odinger equation with the quadratic potential $V(x)$ in eq.~\eqref{eq:FQ2}. 
This covers a free particle for $b=c=0$, or a particle in a harmonic potential for a matrix $c$ with positive eigenvalues. 
The constant $a$ drops out in eq.~\eqref{eq:11}. 
This corresponds to an arbitrary constant shift of the quantum energy.

\subsection*{Free particle}

\indent For a free particle the momentum operator $\hat p=-i\partial_x$ corresponds to a conserved statistical observable,
\begin{equation}
[H,\hat p_k]=0\,.
\label{eq:FQ4}
\end{equation}
The eigenfunctions,
\begin{equation}
\psi_p(x)=L^{-D/2}\exp(ipx)\,,
\label{eq:FQ5}
\end{equation}
are periodic. 
(We may take a $D$-dimensional torus with discrete values of $p$ and take the limit $L\to\infty$ at the end.) 
The momentum observable measures the periodicity in space of the wave function and associated probability distribution. 
The direct product wave function associated to $\psi_p$ is given by
\begin{align}
&\varphi_p
=
L^{-D}\exp\bigl(ip(x-y)\bigr)
=
L^{-D}\exp(ips)\,,
\nn \\
&q_p
=
L^{-D}\delta\,(v-p)\,.
\label{eq:FQ5A}
\end{align}
The probability distribution is homogenous in space with sharp momentum.

\indent For a free particle the general solution of the
Schr\"odinger equation for $\psi(x)$ is a superposition of the basis
functions \eqref{eq:FQ5}, with time dependence of each mode given by the frequency $\omega(p)=\frac{p^2}{2}$.
This does not correspond to a superposition of $\varphi_p$ or $q_p$.
For example, the real wave function
\begin{equation}
\psi
=
\frac{1}{\sqrt{2}}L^{-D/2}
\left(
e^{ipx}+e^{-ipx}
\right)
=
\sqrt{2}\,L^{-D/2}\cos(px)
\label{eq:FQ5B}
\end{equation}
corresponds to
\begin{equation}
q
=
L^{-D}
\left[
\cos(2pz)
+
\frac{1}{2}
\bigl(
\delta(v-p)+\delta(v+p)
\bigr)
\right]\,.
\label{eq:FQ5C}
\end{equation}
The corresponding probability distribution $w=q^2$ is already rather complex.

\subsection*{Quantum dispersion}

\indent The time evolution of a localized quantum particle always shows the phenomenon of dispersion. 
This is not necessarily the case for a probabilistic classical particle. 
It is instructive to see how the general quantum property of dispersion emerges from the classical probabilistic setting. 
Let us consider one space dimension and a Gaussian wave packet
\begin{equation}
\psi(t,x)
=
\int_p
g(p)\,
\exp\left[
ip(x-x_0)-i\frac{p^2}{2}t
\right]\,,
\label{eq:FQ6}
\end{equation}
with
\begin{equation}
g(p)
=
\left(\frac{2\pi}{\Delta_p}\right)^{1/4}
\exp\left(
-\frac{(p-p_0)^2}{4\Delta_p}
\right)\,,
\label{eq:FQ7}
\end{equation}
obeying
\begin{equation}
\int_p g^2(p)=1\,,
\quad
\int_x \psi^*(x)\psi(x)=1\,.
\label{eq:FQ8}
\end{equation}
The Fourier transform of $\psi(t,x)$ reads
\begin{align}
\tilde\psi(t,p)
&=
\int_x e^{-ipx}\psi(t,x)
\nn \\
&=
g(p)
\exp\left[
-i\left(px_0+\frac{p^2}{2}t\right)
\right]\,,
\label{eq:FQ9}
\end{align}
and one has
\begin{equation}
\langle p\rangle=p_0\,,
\quad
\left\langle (p-p_0)^2\right\rangle=\Delta_p\,.
\label{eq:FQ10}
\end{equation}
The relations \eqref{eq:FQ10} do not depend on time since $p$ is a conserved quantity. 

\indent Performing the $p$-integration, the wave packet \eqref{eq:FQ6} becomes
\begin{align}
&\psi(t,x)
=
\left(\frac{2\Delta_p}{\pi}\right)^{\frac{1}{4}}
\bigl(1+2i\Delta_p t\bigr)^{-\frac{1}{2}}
\exp\tilde A\,,
\nn \\
&\tilde A
=
\frac{
ip_0\left(x-x_0-\frac{p_0}{2}t\right)
-\Delta_p(x-x_0)^2
}{
1+2i\Delta_p t
}\,,
\label{eq:FQ11}
\end{align}
with probability to find the particle at the quantum position $x$
\begin{equation}
|\psi(t,x)|^2
=
\bigl(2\pi\Delta_x(t)\bigr)^{-\frac{1}{2}}
\exp\left(
-\frac{(x-\bar x(t))^2}{2\Delta_x(t)}
\right)\,,
\label{eq:FQ12}
\end{equation}
where
\begin{align}
&\bar x(t)=\langle x\rangle=x_0+p_0t\,,
\nn \\
&\Delta_x(t)
=
\left\langle (x-\bar x)^2\right\rangle
=
\frac{1}{4\Delta_p}+\Delta_p t^2\,.
\label{eq:FQ13}
\end{align}
The wave function broadens according to the increase of $\Delta_x$ for $t\to\infty$.

\subsection*{Classical dispersion}

\indent In contrast, we may consider a classical wave packet at $t=0$,
\begin{align}
&q(z,v)=\bar g(v)\,\bar f(z)\,,
\nn \\
&\bar g(v)
=
\left(\frac{2\pi}{\Delta_v}\right)^{\frac{1}{4}}
\exp\left(
-\frac{(v-v_0)^2}{4\Delta_v}
\right)\,,
\nn \\
&\bar f(z)
=
\bigl(2\pi\bar\Delta_z\bigr)^{-\frac{1}{4}}
\exp\left(
-\frac{(z-z_0)^2}{4\bar\Delta_z}
\right)\,.
\label{eq:FQ14}
\end{align}
The Liouville equation for the free particle,
\begin{equation}
(\partial_t+v\partial_z)\,q=0\,,
\label{eq:FQ15}
\end{equation}
has the general solution
\begin{equation}
q(t,z,v)=q(z-vt,v)\,.
\label{eq:FQ16}
\end{equation}
This leaves $\bar g(v)$ untouched, while $\bar f(z)$ is replaced by
\begin{equation}
\bar f(t,z,v)=\bar f(z-vt)\,.
\label{eq:FQ17}
\end{equation}

\indent The expectation values of functions of $v$ are time independent
\begin{equation}
\langle v\rangle=v_0\,,
\quad
\left\langle (v-v_0)^2\right\rangle=\Delta_v\,.
\label{eq:FQ18}
\end{equation}
On the other hand, the expectation value of $z$ shows dispersion if $\Delta_v\neq0$,
\begin{align}
&\bar z(t)=\langle z\rangle=z_0+v_0t\,,
\nn \\
&\Delta_z(t)
=
\left\langle (z-\bar z)^2\right\rangle
=
\bar\Delta_z+\Delta_v \,t^2\,.
\label{eq:FQ19}
\end{align}
This can be understood by employing eqs.~\eqref{eq:22}\eqref{eq:23},
\begin{align}
\partial_t\langle \hat z\rangle
&=
\left\langle i[H,\hat z]\right\rangle
=
\langle \hat v\rangle
=
v_0\,,
\nonumber\\
\partial_t\langle \hat z^2\rangle
&=
\left\langle i[H,\hat z^2]\right\rangle
=
\langle 2\hat v\hat z\rangle\,,
\nonumber\\
\partial_t\langle \hat v\hat z\rangle
&=
\left\langle i[H,\hat v\hat z]\right\rangle
=
\langle \hat v^2\rangle
=
v_0^2+\Delta_v\,.
\label{eq:FQ20}
\end{align}
The dispersion in $\Delta_z$ vanishes for $\Delta_v\to0$. 
This is compatible with arbitrary $\Delta_z(0)$, including the limit $\Delta_z(0)\to0$ which describes a classical point particle. 
In contrast, for the limit $\Delta_p\to0$ in eq.~\eqref{eq:FQ13} one has $\Delta_x\to\infty$. 
This limit corresponds to the plane wave \eqref{eq:FQ5}, which describes no longer a localized particle.

\subsection*{Quantum particle as classical wave packet}

\indent One may ask how the quantum particle is related to classical wave packets.
For this purpose we express the classical wave packet in terms of the quantum observables.
We will then see that the ``pure state condition'' \eqref{eq:FQ3} implies a constraint for the
Gaussian classical wave packet.
If this constraint is not met, one deals with a more general density matrix for the quantum particle.

\indent For the wave function with initial condition \eqref{eq:FQ14} we can perform the Fourier transform \eqref{eq:4}
\begin{align}
&\varphi(t;z,s)
=
\int_v e^{isv}\,\bar g(v)\,\bar f(z-vt)
\nn \\
&=
(\bar\Delta_z\Delta_v)^{-\frac{1}{4}}
\exp\left(
-\frac{(z-z_0)^2}{4\bar\Delta_z}
\right)
\int_v \exp\bigl(-B(v)\bigr)\,,
\label{eq:FQ21}
\end{align}
with
\begin{equation}
B(v)
=
\frac{(v-v_0)^2}{4\Delta_v}
+
\frac{v^2t^2}{4\bar\Delta_z}
-
\frac{(z-z_0)vt}{2\bar\Delta_z}
-
isv\,.
\label{eq:FQ22}
\end{equation}
This yields
\begin{align}
&\varphi(t;z,s)
=
\bigl(\pi\Delta_z(t)\bigr)^{-\frac{1}{2}}
(\Delta_v\bar\Delta_z)^{\frac{1}{4}}
\exp\left\{
-\frac{C}{\Delta_z(t)}
\right\}\,,
\nn \\
&C
=
\frac{1}{4}\bigl(z-\bar z(t)\bigr)^2
+
\Delta_v\bar\Delta_z\,s^2
\nn \\
&\hspace{0.5cm}-
is\bigl[(z-\bar z(t))\Delta_v t+v_0\Delta_z(t)\bigr]\,.
\label{eq:FQ23}
\end{align}
Switching to the variables $x$ and $y$, with
\begin{equation}
\tilde x=x-\bar z(t)\,,
\quad
\tilde y=y-\bar z(t)\,,
\quad
C=C_x+C_y+C_{xy}\,,
\label{eq:FQ24}
\end{equation}
one obtains
\begin{align}
C_x
&=
\left(
\frac{1}{16}
+
\Delta_v\bar\Delta_z
-
\frac{i}{2}\Delta_v t
\right)\tilde x^2
-
iv_0\Delta_z(t)\tilde x\,,
\nonumber\\
C_y
&=
\left(
\frac{1}{16}
+
\Delta_v\bar\Delta_z
+
\frac{i}{2}\Delta_v t
\right)\tilde y^2
+
iv_0\Delta_z(t)\tilde y\,,
\nonumber\\
C_{xy}
&=
2\left(
\frac{1}{16}-\Delta_v\bar\Delta_z
\right)\tilde x\tilde y\,.
\label{eq:FQ25}
\end{align}

\indent In general, the wave function $\varphi(t,x,y)$ is not of the direct product form \eqref{eq:FQ3} due to the term involving $C_{xy}$. 
The direct product form is only realized for the particular relation
\begin{equation}
\Delta_v\bar\Delta_z=\frac{1}{16}\,,
\label{eq:FQ26}
\end{equation}
which yields
\begin{equation}
\psi(x)
=
\bigl(4\pi\Delta_z(t)\bigr)^{-\frac{1}{4}}
\exp\left\{
-
\frac{(1-4i\Delta_v t)\tilde x^2}{8\Delta_z(t)}
\right\}
\exp(iv_0\tilde x)\,.
\label{eq:FQ27}
\end{equation}
For $x_0=z_0$, $p_0=v_0$ and
\begin{equation}
\Delta_x(t)=2\Delta_z(t)\,,
\quad
\Delta_p=2\Delta_v\,,
\label{eq:FQ28}
\end{equation}
this agrees with eq.~\eqref{eq:FQ11}. 
This may be shown by writing
\begin{equation}
\psi(x)=|\psi(x)|e^{i\alpha(x)}\,,
\quad
\alpha
=
v_0\tilde x
+
\frac{\Delta_p t\,\tilde x^2}{2\Delta_x(t)}\,.
\label{eq:FQ29}
\end{equation}
The agreement of $|\psi(x)|$ with eq.~\eqref{eq:FQ12} is seen immediately. 
The phase agrees at $t=0$, and both expressions eqs.~\eqref{eq:FQ27} and \eqref{eq:FQ11} obey the same Schr\"odinger equation for their time evolution. 
Thus the phase is the same for all $t$.
In conclusion, the pure state wave packet for the quantum particle obtains as a Gaussian wave packet for the classical probabilistic particle with the constraint \eqref{eq:FQ26}.

\subsection*{Density matrix}

\indent The condition \eqref{eq:FQ26} may seem to restrict the use of the subsystem for the quantum particle. 
It is, however, only the condition for realizing a pure state for this quantum subsystem, given by eq.~\eqref{eq:FQ3}. 
For a linear force the validity of the subsystem for the quantum particle is universal in the sense that arbitrary classical wave functions project to the density matrix for the quantum subsystem.
The time evolution of the subsystem is closed. 
All observables constructed from the quantum position $x$ and quantum momentum $p$ are represented by operators acting in the subsystem. 
Their expectation values can be computed from the density matrix by the usual quantum rule.

\indent For an arbitrary complex wave function $\varphi(x,y)$ for the probabilistic classical particle the "classical density matrix" is defined by
\begin{equation}
\tilde\rho(x,y;x',y')
=
\varphi(x,y)\varphi^*(x',y')\,.
\label{eq:FQ31}
\end{equation}
The density matrix for the subsystem obtains by "coarse graining" realized by integration over $y$ \cite{CWQP2},
\begin{equation}
\rho(x,x')
=
\int_y \tilde\rho(x,y;x',y)\,.
\label{eq:FQ32}
\end{equation}
The time evolution for $\rho$ obeys the von Neumann equation with the Hamiltonian $H^{(p)}$ for the quantum particle,
\begin{equation}
i\partial_t\rho
=
\bigl[H^{(p)},\rho\bigr]\,.
\label{eq:FQ33}
\end{equation}
This follows from the decomposition \eqref{eq:11} for $H^{(\mathrm{int})}=0$.
The expectation values of general observables are given in terms of the classical density matrix $\tilde\rho$ by the standard quantum rule
\begin{equation}
\langle A\rangle
=
\varphi^\dagger\hat A\varphi
=
\operatorname{tr}(\hat A\tilde\rho)\,.
\label{eq:FQ34}
\end{equation}
Observables built from $\hat x$ and $\hat p$ act on $x$, but not on $y$. 
For these quantum observables the expectation value is given by the reduced density matrix \eqref{eq:FQ32},
\begin{equation}
\langle A\rangle
=
\operatorname{tr}(\hat A\rho)\,.
\label{eq:FQ35}
\end{equation}
The relations \eqref{eq:FQ31}--\eqref{eq:FQ35} can be transformed to an arbitrary basis. 
For a linear force the probabilistic description of a classical particle always admits a quantum subsystem for which the Hamiltonian is the usual quantum Hamiltonian, the probabilistic information is encoded in a density matrix, and all laws of quantum mechanics apply. 
This holds provided the observables are restricted to the quantum observables constructed from $\hat x$ and $\hat p$.

\indent As an example we may compute the reduced density matrix for an arbitrary classical Gaussian wave packet \eqref{eq:FQ14} at $t=0$, for which
\begin{align}
\label{eq:FQ36}
&\varphi(0;x,y)
=
\left(
\frac{\Delta_v}{\pi^2\Delta_z}
\right)^{\frac{1}{4}}
\\
&\times\exp\Bigg\{
-
\Bigg[
\frac{1}{16\Delta_z}(\tilde x+\tilde y)^2
+
\Delta_v(\tilde x-\tilde y)^2
-
iv_0(\tilde x-\tilde y)
\Bigg]
\Bigg\},\nn
\end{align}
For the reduced density matrix \eqref{eq:FQ32} one obtains
\begin{align}
&\rho(x,x')
=
\left(
\frac{8\Delta_v}{\pi}
\right)^{\frac{1}{2}}
\bigl(1+16\Delta_v\Delta_z\bigr)^{-\frac{1}{2}}
\exp\bigl(iv_0(\tilde x-\tilde x')\bigr)
\nonumber\\
&\quad\times
\exp\left\{
-
\left(
\frac{1}{16\Delta_z}+\Delta_v
\right)
(\tilde x^2+\tilde x'^2)
\right\}
\\
&\quad\times
\exp\left\{
\left(
\frac{1}{8\Delta_z}+2\Delta_v
\right)^{-1}
\left(
\frac{1}{16\Delta_z}-\Delta_v
\right)^2
(\tilde x+\tilde x')^2
\right\}\,. \nonumber
\label{eq:FQ37}
\end{align}
If the condition \eqref{eq:FQ26} holds, the last exponential equals one and $\rho(x,x')$ is a pure state density matrix, $\rho(x,x')=\psi(x)\psi^*(x')$.

\subsection*{Quantum harmonic oscillator}

\indent For a general linear force \eqref{eq:FQ1} we may eliminate $b_k$ by a constant shift in $z_k$. 
Without loss of generality we may set $a=0$ for the potential \eqref{eq:FQ2}. 
For the remaining quadratic potential we assume that the matrix $c$ has only positive eigenvalues. 
We can then diagonalize $c$ by an orthogonal transformation of $z_k$.
Transforming $v_k$ with the same orthogonal transformation the Liouville equation remains unchanged. 
One ends with a number of decoupled oscillators. 
We can therefore restrict the discussion to the one-dimensional case, with quadratic potential
\begin{equation}
V=\frac{c}{2}z^2\,,
\quad
c>0\,.
\label{eq:FQ38}
\end{equation}
Here $c$ corresponds to one of the eigenvalues of the matrix $c_{kl}$.

\indent Consider first a product wave function \eqref{eq:FQ3}. 
It corresponds to a pure state for the reduced quantum system, with wave function obeying the Schr\"odinger equation
\begin{equation}
i\partial_t\psi
=
H^{(p)}\psi
=
-\frac{1}{2}\partial_x^2\psi
+
\frac{\omega^2}{2}x^2\psi
\,,
\quad
\omega^2=c\,.
\label{eq:FQ39}
\end{equation}
This is the standard quantum harmonic oscillator. 
The eigenvalues of the quantum Hamiltonian $H^{(p)}$ are discrete and equidistant, with positive integer $n$,
\begin{equation}
H^{(p)}\psi_n(x)=E_n\psi_n(x)\,,
\quad
E_n=\left(n+\frac{1}{2}\right)\omega\,.
\label{eq:FQ40}
\end{equation}
The eigenfunctions of the quantum energy form a complete orthonormal basis,
\begin{equation}
\int_x \psi_n(x)\psi_m^*(x)=\delta_{nm}\,.
\label{eq:FQ40A}
\end{equation}
The time evolution for an arbitrary wave function reads
\begin{equation}
\psi(t,x)
=
\sum_n a_n\,\psi_n(x)\,e^{-iE_n t}\,.
\label{eq:FQ41}
\end{equation}
Quantum observables are represented by operators which are functions of $\hat x$ and $\hat p$. 
Their expectation values coincide with the well-known results for the quantum harmonic oscillator.

\indent A general complex wave function for the probabilistic classical particle can be written as a double expansion in eigenfunctions $\psi_n$,
\begin{equation}
\varphi(t;x,y)
=
b_{mn}(t)\,
\psi_m(x)\psi_n^*(y)\,.
\label{eq:FQ42}
\end{equation}
The selection rule \eqref{eq:10} requires the coefficient $b_{mn}$ to be an element of a hermitian matrix,
\begin{equation}
b^\dagger=b\,,
\quad
b_{mn}^*=b_{nm}\,,
\quad
\text{tr}\,b^2=1\,,
\label{eq:FQ43}
\end{equation}
where the last relation follows from the normalization of $\varphi$. 
The density matrix for the reduced quantum system \eqref{eq:FQ32} obtains as
\begin{equation}
\rho(t; x,x')
=
b^2_{mm^\prime}(t)\,
\psi_m(x)\psi_{m^\prime}^*(x^\prime)\,,
\label{eq:FQ44}
\end{equation}
with $b^2_{mm^\prime}$ the matrix elements of the matrix $b^2$,
\begin{equation}
b^2_{mm^\prime} = \sum_n b_{mn} b_{nm^\prime}\,.
\label{eq:FQ45}
\end{equation}
The eigenvalues of $b^2$ are all positive. 
By a unitary transformation one can diagonalize $b^2$. 
This shows that $\rho(t,x,x')$ can be represented as a sum of pure-state density matrices with positive and normalized weights. 
This reasoning extends to the most general
classical wave function and reduced density matrix for a general linear force \eqref{eq:FQ1}. 
In this way a large family of reduced quantum systems with a discrete frequency spectrum can be realized by classical probabilistic particles, or classical probabilistic fields.

\indent It is instructive to compute for the ground state wave function for the quantum harmonic oscillator, namely
\begin{align}
&\psi^{(0)}(t,x)
=
\left(\frac{\omega}{\pi}\right)^{\frac{1}{4}}
\exp\left(-\frac{\omega}{2}x^2\right)
\exp\left(-i\omega t\right)\,,
\nn \\
&H^{(p)}\psi^{(0)} (t,x)
=
\frac{\omega}{2}\psi^{(0)}(t,x)\,,
\label{eq:FQ46}
\end{align}
the associated wave function \eqref{eq:FQ3},
\begin{align}
\varphi^{(0)}
&=
\left(\frac{\omega}{\pi}\right)^{\frac{1}{2}}
\exp\left(-\frac{\omega}{2}(x^2+y^2)\right)
\nn \\
&=
\left(\frac{\omega}{\pi}\right)^{\frac{1}{2}}
\exp\left(-\omega\left(z^2+\frac{s^2}{4}\right)\right)\,.
\label{eq:104A}
\end{align}
The corresponding real classical wave function takes a simple form
\begin{equation}
q^{(0)}
=
2\exp(-\omega z^2)\exp\left(-\frac{v^2}{\omega}\right)
=
2\exp\left(-\frac{2}{\omega}E_{\mathrm{cl}}\right)\,.
\label{eq:FQ47}
\end{equation}
and this extends to the probability density.
The time-dependence has dropped out.

\indent The eigenfunction for the next level of the quantum energy, $E_1=3\omega/2$, reads similarly
\begin{equation}
\psi^{(1)}
=
\left(\frac{4\omega^3}{\pi}\right)^{1/4}
x\exp\left(-\frac{\omega}{2}x^2\right)
\exp\left(-\frac{3i\omega t}{2}\right)\,.
\label{eq:FQ47A}
\end{equation}
The corresponding classical wave function is given by
\begin{equation}
q^{(1)}
=
\left(\frac{4}{\omega}E_{\mathrm{cl}}-1\right)\,q^{(0)}\,.
\label{eq:FQ48}
\end{equation}
The classical probability distributions for the eigenstates of $H^{(p)}$ are static. 
Oscillating wave functions obtain for superpositions of eigenfunctions with different $E_n$. 
We will discuss this in sect.~\ref{sec:XI}.

\subsection*{Free particles with boundaries}

\indent Consider the probability distribution for a free classical particle in a region of flat space with boundaries. 
In practice this can be realized by excluding the particle from "forbidden regions", for example
by imposing an arbitrarily large potential $V_f$ for the forbidden region, while maintaining $V=0$ in the allowed region. 
In this way one can describe walls or more complex geometries as walls with holes. 
Of particular interest is the question how the probabilistic classical particle behaves in the double-slit situation for which quantum mechanics shows the characteristic interference effects.

\indent As a start we discuss the one-dimensional particle with a wall at $z_b$. 
The region $z<z_b$ is allowed, while the region $z>z_b$ is forbidden. 
For the allowed region the Liouville equation for the free particle applies.
We need the boundary condition at $z_b$.
It is given by the requirement of zero probability for the particle to enter the forbidden region
\begin{equation}
w(z_b,v)\,v=0\,.
\label{eq:PB1}
\end{equation}
If this condition would be violated one would have a non-zero probability for a particle at $z_b$ with momentum or velocity $v$. 
For $v>0$ the particle would enter the forbidden region, and for $v<0$ it would come from the forbidden region. 
Both cases are forbidden. 

\indent With $w=q^2$, the condition \eqref{eq:PB1} translates directly to the boundary condition for the wave function
\begin{equation}
q(z_b,v)\,v=0\,.
\label{eq:PB2}
\end{equation}
Since this has to hold for all $v$ we conclude that at $z_b$ the wave function is of the form
\begin{equation}
q(z_b,v)=\tilde q\,(z_b)\,\delta(v)\,.
\label{eq:PB3}
\end{equation}
By the Fourier transform this translates to the complex wave function
\begin{equation}
\varphi(z_b,s)=\tilde\varphi(z_b)\,.
\label{eq:PB4}
\end{equation}
In other words, at the boundary $z_b$ the complex wave function is independent of $s$ or $x-y$. 
The condition \eqref{eq:PB4} can be written in differential form as
\begin{equation}
\partial_s\varphi(z_b,s)=0\,.
\label{eq:PB5}
\end{equation}
For a pure state of the reduced quantum system \eqref{eq:FQ3} the condition \eqref{eq:PB5} reads
\begin{equation}
\partial_s
\left[
\psi\!\left(z_b+\frac{s}{2}\right)
\psi^*\!\left(z_b-\frac{s}{2}\right)
\right]
=0\,.
\label{eq:PB6}
\end{equation}
This entails
\begin{equation}
\frac{\partial\psi}{\partial x}
\!\left(z_b+\frac{s}{2}\right)
\psi^*\!\left(z_b-\frac{s}{2}\right)
=
\psi\!\left(z_b+\frac{s}{2}\right)
\frac{\partial\psi^*}{\partial x}
\!\left(z_b-\frac{s}{2}\right)\,,
\label{eq:PB7}
\end{equation}
which has to hold for all $s$.

\indent One simple realization of the boundary condition \eqref{eq:PB2} is the Dirichlet boundary condition
\begin{equation}
q(z_b,v)=0\,,
\quad
\varphi(z_b,s)=0\,.
\label{eq:PB8}
\end{equation}
For a pure state of the reduced quantum system this entails for all $s$
\begin{equation}
\psi\!\left(z_b+\frac{s}{2}\right)
\psi^*\!\left(z_b-\frac{s}{2}\right)
=0\,.
\label{eq:PB9}
\end{equation}
We can fulfil this requirement by
\begin{align}
\psi\!\left(z_b+\frac{s}{2}\right)&=0
\quad\text{for }s\ge0
\nn \\
\psi^*\!\left(z_b-\frac{s}{2}\right)&=0
\quad\text{for }s\le0\,.
\label{eq:PB10}
\end{align}
These conditions combine to
\begin{equation}
\psi(x\ge z_b)=0\,.
\label{eq:PB11}
\end{equation}
The pure-state wave function $\psi(x)$ vanishes for all $x$ in the forbidden region, including the boundary at $x=z_b$. 
The boundary condition \eqref{eq:PB10} or\eqref{eq:PB11} is consistent with the more general boundary condition \eqref{eq:PB7}. 
Both sides vanish since one of the factors in each product vanishes.

\indent In the multidimensional case $z_b$ denotes locations on a hypersurface which constitutes the boundary. 
In eq.~\eqref{eq:PB2} $v$ is the component of the momentum or velocity normal to the hypersurface. 
The Dirichlet boundary condition \eqref{eq:PB8} generalizes for a pure state of the reduced quantum system to $\psi(x\in F)=0$, if $F$ is the forbidden region with the boundary included. 
We observe that the condition $\psi(x\in F)=0$ does not require $q(x_B)=0$. 
It may be realized with the more general boundary condition \eqref{eq:PB4}. 
One may also encounter more general boundary conditions for the wave function of a pure quantum state for the quantum particle.

\indent In general, it is not guaranteed that the physics at the boundary
respects the factorization into subsystems for the quantum particle and the mirror particle. 
Boundaries could  induce an effective mixing between the quantum particle and the mirror particle. 
Such an effect would not be compatible with a unitary evolution of the subsystem for the quantum particle. 
For a consistent treatment of the quantum particle in a bounded geometry, it is important that the unitarity of the time evolution for the subsystem is preserved, at least to a good approximation.

\subsection*{Double-slit experiment}

\indent Can the probabilistic classical particle produce the interference
pattern of the double-slit experiment which is characteristic for quantum mechanics? 
The conceptual answer is yes. 
It needs, however, particular initial conditions and specifications. 
The general phenomenon of interference is characteristic for our setting. 
The superposition principle for solutions of the Schr\"odinger equation holds for the
wave function $\psi$, as well as for $\varphi$ and $q$. 
The probabilities are quadratic in $\varphi$ and $q$, and even quartic in $\psi$. 
This is the key property for linear waves and interference. 
As for quantum mechanics the wave aspect concerns only the probabilistic information. 
One can send single well separated particles to the slits. 
The interference pattern concerns the probability to detect a particle at a certain location on the screen behind the slits.

\indent The double-slit experiment is characterized by a particular geometry for the
propagation of free particles. This concerns a wall (forbidden
region) at $z_2=0$, with two open slits centered at $z_1=-d/2$ and $z_1=d/2$. 
The slits may have a certain width, and in practice the wall may have a certain thickness. 
The geometry is independent of $z_3$. 
This setup defines the forbidden region.
One has to implement the boundary condition that for all points of the forbidden region the probability to find the particle vanishes.
For the allowed region the Liouville equation for free classical particles is valid.
We have already seen how the corresponding Schr\"odinger equation for the subsystem of the quantum particle follows.
Possible obstructions for the realisation of the double-slit interference can only arise from the boundary conditions.

\indent For the Schr\"odinger equation for a free quantum particle we know that
solutions with appropriate boundary conditions exist.
They show the interference pattern well known from quantum physics. 
For such a solution we can construct $\varphi(x,y)$ or $\varphi(z,s)$ according to
eq.~\eqref{eq:FQ3}, and $q(z,v)$ by inverting the Fourier transform \eqref{eq:4}. The real
wave function $q(z,v)$ is then a solution of the Liouville equation for
the free classical particle. This extends to the probability
distribution $w=q^2$. This procedure proves that a classical probability
distribution exists which solves the Liouville equation and produces the
interference pattern for observables based on the quantum position $x$.
The issue is therefore not so much the existence of classical
probabilities realising the interference of the double-slit experiment,
but rather if such solutions require (perhaps extreme) tuning of boundary physics or initial conditions.
In quantum mechanics the double-slit interference is robust with respect to small changes of the detailed
boundary conditions or initial conditions.
The question arises if the quantum-mechanical solution for $\psi$ can be realized by a choice of boundary
potential and initial value without extreme tuning.

\indent For a realisation without tuning the physical realisation of the wall should be compatible with the quantum subsystem in the sense
that the time evolution of the subsystem wave function $\psi$ or
density matrix $\rho$ remains closed and unitary. We may require the
Dirichlet boundary condition $\psi(z_b) = 0$
at the boundary between the allowed and forbidden region. 
The time evolution of the probabilistic information for the classical
particle is then described by the Liouville equation for a free particle
with a Dirichlet boundary condition for $\psi$. 

\indent The next step is the preparation of a wave packet which corresponds to a pure state for the
subsystem describing the quantum particle. This needs to implement the pure state condition similar to  eq.~\eqref{eq:FQ26} for
the classical wave packet \eqref{eq:FQ14}. To a good approximation the
three-dimensional classical wave packet does not depend on $z_1$ or
$z_3$, such that $z$ in the Gaussian $\bar f (z)$ in eq.~\eqref{eq:FQ14} stands for $z_2$. 
At $t=0$ we choose a large negative value of $z_2$. 
The initial wave packet starts far from the wall, such that
for time near $t=0$ the wall plays no role. For $\bar g(v)$ in
eq.~\eqref{eq:11} we take $v=v_2$, with positive $v_0$, such that the
wave packet will reach the wall at some later time. For the distribution in
$v_1$ and $v_3$ we take $\delta$-functions $\delta(v_1) \delta(v_3)$. 
(In practice one may allow a small spread in $v_1$ and $v_2$, and limited extension in $z_1$ and $z_3$.)

\indent Formally, the precise specification of all details of the wave function $q$ at $t=0$ determines the system
completely, since the Liouville equation is a first order differential
equation. Not every initial $q(t=0)$ will be compatible with some precise boundary conditions. 
As long as the physical realization of a boundary which is consistent with the unitarity of the subsystem
does not need fine-tuning, this poses no problem since boundary conditions have to hold only approximately. 
In this case one may prefer to work with exact boundary conditions and impose mild restrictions
on the precise choice of the initial wave function in order to implement formal consistency.
This will not affect the dynamics to a good approximation.

\indent With given initial conditions the dynamics of the probabilistic
information is determined by the Liouville equation for the free
classical particle in a restricted geometry. Provided that the boundary
physics preserves the unitarity of the evolution for the quantum subsystem this yields
the von Neumann equation for the free quantum particle in the restricted geometry.
With the pure state condition it translates directly to the evolution of $\psi(t,x)$
according to the Schr\"odinger equation for a free particle in the
restricted geometry. One finds precisely the setting for the quantum wave function
in the two-slit experiment. The well known results take over.
Behind the screen at some positive $z_2$ one finds the characteristic interference pattern
for the dependence of $\psi^\ast (x) \psi(x)$ on $x_1$.
For a pronounced interference pattern one needs small $\Delta_v$,
the dispersion in momentum or velocity diminishes the interference effects.

\indent Under these conditions the classical probabilistic particle realises the quantum
interference of the double-slit experiment. An important ingredient is the pure state condition \eqref{eq:FQ26}.
For deviations from this condition the quantum particle is described by a density matrix with a tendency to wash out interference effects. 
A crucial requirement is the (approximate) consistency of the boundary
physics with the unitary evolution of the reduced quantum system.
Deviations from a unitary evolution are expected to diminish the
interference pattern. If the deviation from subsystem-unitarity is
substantial, one may need fine tuning of initial conditions in order to implement the interference pattern.

\indent It is possible to realize the interference pattern for a more
macroscopic situation for classical particles? The necessary geometry in "physical
units" obtains from the setting with $m=1$ by appropriate rescalings. 
The challenge is to prepare the boundary condition for $\psi(x)$ consistent with its unitary evolution, and to prepare the pure state wave packet. 
One also needs to understand what the detectors behind the screen measure in practice.
Does a hit in a detector correspond to a particle at $z$, or rather to a "quantum particle" at $x$?

\section{Quantum angular momentum}
\label{sec:VI}

\indent For a classical particle in three dimensions moving in a potential with rotation symmetry the quantum angular momentum is a conserved statistical observable. 
The operators for its components obey the usual commutation relation in quantum theory. 
They provide a further simple example for observables whose associated operators do not commute.

\indent Consider a potential $V(r)$ with associated force $F_k$,
\begin{equation}
F_k = - \frac{\partial}{\partial z_k} V(r)\,,
\quad
r^2 = \sum_{k=1}^3 z_k^2 \,.
\label{eq:AM1}
\end{equation}
The classical angular momentum 
\begin{equation}
    \label{eq:AM1A}
    L_k^{(\text{cl})} = \varepsilon_{klm} z_l v_m
\end{equation}
is a conserved quantity.
Its associated operator $\hat L_k^{(\text{cl})}$ is diagonal in the basis $q(t)$. 
It commutes with the Hamiltonian, and its different components commute
\begin{equation}
[H,\hat L_k^{(\text{cl})}] = 0\,,
\quad
[\hat L_k^{(\text{cl})},\hat L_l^{(\text{cl})}] = 0 \,.
\label{eq:AM2}
\end{equation}
Wave functions which depend only on $E_{cl}$ and $L_k^{(\text{cl})}$ are static, with associated static probability distributions.
In a microphysical situation with substantial dispersion of velocities and positions the classical angular momentum is not a very suitable observable.
Its measurement would require sharp values for both position and momentum.

\indent A more robust statistical observable is the quantum angular momentum $L_k$, with associated operator
\begin{align}
\hat L_k
&= - i \varepsilon_{klm}
\left(
z_l \frac{\partial}{\partial z_m}
+
v_l \frac{\partial}{\partial v_m}
\right)
\nn \\
&= - i \varepsilon_{klm}
\left(
z_l \frac{\partial}{\partial z_m}
+
s_l \frac{\partial}{\partial s_m}
\right)
\nn \\
&= - i \varepsilon_{klm}
\left(
x_l \frac{\partial}{\partial x_m}
+
y_l \frac{\partial}{\partial y_m}
\right) \,.
\label{eq:AM3}
\end{align}
It is the generator for infinitesimal rotations and obeys the standard commutation relation
\begin{equation}
[\hat L_k,\hat L_l] = i \varepsilon_{klm}\hat L_m \,.
\label{eq:AM4}
\end{equation}
As familiar in quantum mechanics, rotation symmetry induces a conserved quantum angular momentum,
\begin{equation}
[H,\hat L_k] = 0 \,.
\label{eq:AM5}
\end{equation}
The quantum angular momentum differs from the classical angular momentum.
The spectrum of eigenvalues of the classical angular momentum
is continuous. In contrast, the discrete spectrum of the quantum angular momentum follows from the commutation relation \eqref{eq:AM4}.

\indent For the particular case of a rotation invariant harmonic potential the quantum particle and the mirror particle decouple. 
One has therefore separately conserved angular momentum for the particle and the mirror particle,
\begin{align}
\hat L_k^{(p)} = - i \varepsilon_{klm}\, x_l \frac{\partial}{\partial x_m}\,,
&\quad
\hat L_k^{(m)} = - i \varepsilon_{klm}\, y_l \frac{\partial}{\partial y_m}\,,
\nn \\
\hat L_k &= \hat L_k^{(p)} + \hat L_k^{(m)}\,.
\label{eq:AM6}
\end{align}
For the reduced quantum system the operator $\hat L_k^{(p)}$ plays the usual role for the quantum harmonic oscillator. 
In general, the interaction part \eqref{eq:15} of $H$ is invariant only under combined rotations of $x$ and $y$.
Then $L_k^{(p)}$ and $L_k^{(m)}$ are not conserved separately.

\indent The quantum angular momentum commutes with the squared
classical squared angular momentum $\hat L_{cl}^{2} = \hat{L}^{(\text{cl})}_k \hat{L}^{(\text{cl})}_k$ and the classical energy,
\begin{equation}
[\hat L_k,\hat L_{cl}^{2}] = 0\,,
\quad
[\hat L_k,\hat E_{cl}] = 0\,.
\label{eq:AM8}
\end{equation}
The components of the classical angular momentum are rotated according to
\begin{equation}
[\hat L_k,\hat L_l^{(\text{cl})}]
=
i\varepsilon_{klm}\hat L_m^{(\text{cl})}\,,
\quad
[\hat L^{2},\hat L_l^{(\text{cl})}]=0\,.
\label{eq:AM9}
\end{equation}
This implies that given components commute,
\begin{equation}
[\hat L_3,\hat L_3^{(\text{cl})}]=0\,,
\quad
[\hat L_k,\hat L_k^{(\text{cl})}]=0\,.
\label{eq:AM10}
\end{equation}
One could have simultaneous sharp values of $\hat L_3$, $\hat L^2$,
$\hat L_3^{(\text{cl})}$, $\hat L_{cl}^{2}$, $\hat E_{cl}$ and $H$.

\section{Eigenstates of quantum angular momentum}
\label{sec:VII}

\indent The classical and quantum angular momentum can have rather different expectation values. 
The expectation value of the classical angular momentum yields the mean angular momentum of the classical particle. 
It differs from zero for a particle rotating around a given axis with a statistically preferred sign of angular velocity. 
In contrast, the expectation value of the quantum angular momentum makes a statement about the properties of the probability distribution with respect to rotations.
For example, an eigenfunction for a zero value of $\hat L_k$ corresponds to a wave function or probability distribution which is invariant under rotations around the $k$-axis. 
This can be realized for probability distributions with an arbitrarily large mean value of the classical angular momentum $\langle L_k^{(\text{cl})}\rangle$.
In order to get some intuition about the quantum angular momentum observable we discuss here a particular eigenstate of $\hat L$.
It could describe a particular ``quantum state'' for a cloud of non-interacting dust particles rotating around a star, or the probability to find a particle in some central potential.

\subsection*{Quantum numbers for classical particles}

\indent The eigenvalues of $\hat L^2$ are $l(l+1)$ with integer $l$.
As an example we discuss eigenfunctions in the vector
representation $l=1$. In the $(z,s)$-basis we consider three
wave functions
\begin{equation}
\psi_k = a(r,u,w)\,z_k + b(r,u,w)\,s_k \,,
\label{eq:AM11}
\end{equation}
with $a$ and $b$ complex functions of the invariants
\begin{equation}
r^2 = z_k z_k\,,
\quad
u^2 = s_k s_k\,,
\quad
w = z_k s_k \,.
\label{eq:AM12}
\end{equation}
The normalization of $\psi_k$ is given by (no sum over $k$ here)
\begin{equation}
\int_{z,s} |\psi_k|^2
=
\int_{z,s}
\Bigl(
|a|^2 z_k^2 + |b|^2 s_k^2 + (ab^*+a^*b)\,z_ks_k
\Bigr)
=1\,.
\label{eq:AM13}
\end{equation}
Rotations are maps in the space of the three wave functions
\begin{equation}
\hat L_l \psi_k = i\varepsilon_{lkm}\psi_m\,,
\quad
\hat L^2 \psi_k = 2\psi_k \,.
\label{eq:AM13A}
\end{equation}
Thus $\psi_k$ are basis functions of a vector representation of the rotation group 
They are eigenfunctions of $\hat{L}_k$ with eigenvalue zero
and therefore invariant under rotations around the $k$-axis.

\indent A general element in this representation is given by
\begin{equation}
\psi = c_k \psi_k\,,
\quad
\sum_k |c_k|^2 = 1\,.
\label{eq:AM13B}
\end{equation}
With $\int_{z,s}\psi_k^*\psi_l=\delta_{kl}$ the normalization \eqref{eq:AM13} 
implies $\int_{z,s} |\psi|^2 = 1$.
A wave function obeying the constraint \eqref{eq:5} can be constructed by
a superposition of $\psi$ with its conjugate wave function
$\bar\psi$ which obtains from $\psi$ by $s\to -s$ and complex conjugation,
\begin{align}
\label{eq:AM13C}
&\bar\psi = c_k^* \bar\psi_k\,,
\quad
\bar\psi_k = \bar a\, z_k - \bar b\, s_k\,, \\
&\bar a(r,u,w)=a^*(r,u,-w)\,,
\quad
\bar b(r,u,w)=b^*(r,u,-w)\,, \nn
\end{align}
namely
\begin{equation}
\varphi
=
\frac{1}{\sqrt{2+d}}(\psi+\bar\psi)\,,
\quad
d=
\int_{z,s}
\left(
\psi^*\bar\psi+\bar\psi^*\psi
\right)\,.
\label{eq:AM13D}
\end{equation}
Instead of a rotation of $\psi_k$ with fixed $c_k$ we can equivalently rotate
the coefficients $c_k$ at fixed $\psi_k$, as determined by the infinitesimal rotation
\begin{align}
\delta\psi
&=
i\alpha_l \hat L_l \psi
=
\delta c_k \psi_k\,,
\nn \\
\delta c_k &= \alpha_l \varepsilon_{lkm} c_m \,.
\label{eq:AM13E}
\end{align}
This extends to $\bar{\psi}$ and $\varphi$.

\indent The functions $a$ and $b$ are invariant under the parity
transformation which switches simultaneously the sign of $z$
and $s$.
Therefore $\psi$ and $\varphi$ are odd under parity.
The expectation values of all parity odd observables vanish.
Those involve only terms with an odd number of factors $z$ or $s$,
e.g. $\langle z_k\rangle = 0$, $\langle z_k z_l z_m\rangle = 0$, etc.
From eq.~\eqref{eq:AM13A} one infers for the quantum angular momentum
\begin{equation}
\langle L_k\rangle = 0\,.
\label{eq:AM13F}
\end{equation}

\indent We may select a wave function $\varphi_3$ with a given "direction",
$|c_3|^2 = 1$, $c_1 = c_2 =0$.
It obeys
\begin{equation}
\hat L_1^{\,2}\varphi_3 = \varphi_3\,,
\quad
\hat L_2^{\,2}\varphi_3 = \varphi_3\,,
\quad
\hat L_3 \varphi_3 = 0\,.
\label{eq:AM13G}
\end{equation}
Thus $\varphi_3$ is an eigenstate of
$\hat L_3$, $\hat L_1^{2}$, $\hat L_2^{2}$.
The direction is singled out by the direction for which the angular momentum takes the sharp value zero.
This property extends to an arbitrary value of the vector $\vec c$ which obtains from $(0,0,c_3)$ by a rotation. 
We can focus our discussion on this type of wave function. 
More general wave functions, as $\vec c=(0,c_2,c_3)$ with real $c_3$ and purely imaginary $c_2$, can be obtained from them by superposition.

\indent The operator for the classical angular momentum reads
\begin{equation}
\hat L_l^{(\text{cl})}
=
- i \varepsilon_{lmn} z_m \frac{\partial}{\partial s_n}\,.
\label{eq:AM13H}
\end{equation}
Not all components can have a sharp value in the state \eqref{eq:AM13G}. 
The non-vanishing commutator \eqref{eq:AM9} implies an uncertainty relation. 
If we choose the state \eqref{eq:AM13G} one could still have a sharp value of $L_3^{(\text{cl})}$, 
since $\hat L_3^{(\text{cl})}$ commutes with $\hat L_3$, $\hat L_1^{2}$ and $\hat L_2^{\,2}$.
If $\psi_3$ is an eigenstate with eigenvalue $\bar L_3$, then
\begin{equation}
\hat L_3^{(\text{cl})}\psi_3
=
- i\left(
z_1\frac{\partial}{\partial s_2}
-
z_2\frac{\partial}{\partial s_1}
\right)\psi_3
=
\bar L_3 \psi_3 \,,
\label{eq:AM13I}
\end{equation}
one finds that $\bar\psi_3$ is an eigenstate with the same
eigenvalue, and therefore $\varphi_3$ is an eigenstate.

\subsection*{Time evolution}

\indent The time evolution of $\psi$ is given by the Hamiltonian \eqref{eq:6}
\begin{equation}
H
=
- \frac{\partial}{\partial z_k} \frac{\partial}{\partial s_k}
+ \frac{w}{r}\frac{\partial V}{\partial r}\,.
\label{eq:AM14}
\end{equation}
One finds
\begin{equation}
H\psi_k = \tilde a\, z_k + \tilde b\, s_k \,,
\label{eq:AM15}
\end{equation}
with
\begin{align}
&\tilde a
=
\frac{w}{r}\frac{\partial V}{\partial r}
-
\Bigg(
4\frac{\partial a}{\partial w}
+\frac{1}{r}\frac{\partial b}{\partial r}
+\frac{w}{ur}\frac{\partial^2 a}{\partial r\,\partial u}
\nn \\
&\hspace{3cm}+u\frac{\partial^2 a}{\partial u\,\partial w}
+r\frac{\partial^2 a}{\partial r\,\partial w}
+w\frac{\partial^2 a}{\partial w^2}
\Bigg)\,,
\nn \\
&\tilde b
=
\frac{w}{r}\frac{\partial V}{\partial r}
-
\Bigg(
4\frac{\partial b}{\partial w}
+\frac{1}{u}\frac{\partial a}{\partial u}
+\frac{w}{ur}\frac{\partial^2 b}{\partial r\,\partial u}
\label{eq:AM16} \\
&\hspace{3cm}+r\frac{\partial^2 b}{\partial r\,\partial w}
+u\frac{\partial^2 b}{\partial u\,\partial w}
+w\frac{\partial^2 b}{\partial w^2}
\Bigg)\,. \nn
\end{align}
Since the evolution is unitary the coefficient functions
$\tilde a$ and $\tilde b$ have to obey the same normalization
condition \eqref{eq:AM13} as $a$ and $b$.
In general, $\tilde a$ and $\tilde b$ differ from $a$ and $b$. 
The time evolution therefore maps a given representation of the rotation group,
characterized by the functions $a$ and $b$, into a different representation.
The coefficients $c_k$ in eq.~\eqref{eq:AM13B} are not changed, however.
An eigenstate with the properties \eqref{eq:AM13G} remains an eigenstate with the same values.
The "direction" of the state is not changed. 
This follows, of course, also directly from $[H,\hat L_3]=0$.
The quantum numbers are preserved during the evolution.

\indent One may investigate possible eigenfunctions of $H$ within this set of wave functions,
\begin{equation}
H\psi_k = \omega \psi_k \,.
\label{eq:AM17}
\end{equation}
For such eigenfunctions $a$ and $b$ have to obey the linear
differential equations
\begin{equation}
\tilde a = \omega a\,,
\quad
\tilde b = \omega b \,.
\label{eq:AM18}
\end{equation}
With $H$ being a real operator which is odd in $s$, our
finds for the conjugate wave function the opposite eigenvalue
\begin{equation}
H\bar\psi_k = -\omega \bar\psi_k \,.
\label{eq:AM19}
\end{equation}
Thus $\varphi_k =(2+d)^{-1/2} (\psi_k+\bar\psi_k)$
is not an eigenfunction of $H$ for $\omega\neq0$.
It is, however, an eigenfunction of $H^2$ with eigenvalue
$\omega^2$. The time evolution of $\varphi_k$ obeys
\begin{equation}
\varphi_k(t)
=
\frac{1}{\sqrt{2+d}}
\left(
e^{-i\omega t}\psi_k(0)
+
e^{i\omega t}\bar\psi_k(0)
\right)\,.
\label{eq:AM20}
\end{equation}
This implies a periodic probability distribution
\begin{align}
\label{eq:AM21}
w(t)&=q^2(t)=\varphi_k^*(t)\varphi_k(t)
\\
&=
\frac{1}{2+d}
\left(
A^2+\bar A^2 + 2A\bar A \cos(2\omega t+\bar\alpha-\alpha)
\right)\,. \nn
\end{align}
Here we employ
\begin{align}
&\psi_k(0)=A e^{i\alpha}\,,
\quad
\bar\psi_k(0)=\bar A e^{i\bar\alpha}\,,
\nn \\
&\int_{z,v} \! A\bar A \cos(\alpha-\bar\alpha)=\frac d2 \,,
\label{eq:AM22}
\end{align}
with real positive $A$, $\bar A$ and real angles $\alpha$, $\bar\alpha$
which depend on $u$, $r$, $w$ according to the solution of the
differential equation \eqref{eq:AM18}.

\indent A general discussion of the solution of the differential
equations \eqref{eq:AM18} may seem to be rather complex. We can infer the
general existence of periodic probability distributions from the
properties of the trajectories of classical point particles. For
any central potential $V(r)$ we deal with an integrable system.
One can find three conserved action variables, with associated
periodic angular variables defining three frequencies $\omega_j$.
Let us select one of the action variables and denote by $\omega^{(\text{cl})}$ 
the frequency associated to it. A periodic probability
distribution $w$ with frequency $\omega^{(\text{cl})}$ can be realized
generically if $w$ does not depend on the angle variables
associated to the two other action variables. Since
$\omega^{(\text{cl})}$ depends on the action variables, $w$ can
have support only for a submanifold of action variables leading to
the same $\omega^{(\text{cl})}$. If we want to identify a periodic
state $\varphi_3$ according to eq.~\eqref{eq:AM13G}
with this type of periodic probability distribution, it has
further to be invariant under rotations around the same axis.
One chooses $L_3^{(\text{cl})}$ as one of the action variables, but
not the one associated to $\omega^{(\text{cl})}$.

\indent A particular case is the Coulomb potential $V(r)=-\kappa/r$.
In this case it is sufficient to fix the classical energy $E^{(\text{cl})}$ of a point particle in order to
obtain a periodic motion with $\omega^{(\text{cl})}$ determined by
$E^{(\text{cl})}$. 
For a search of periodic probability distributions one could focus on eigenstates of the classical energy
\begin{equation}
\hat E^{(\text{cl})} \varphi = E^{(\text{cl})} \varphi \,.
\label{eq:AM23}
\end{equation}
A fixed value of $E^{(\text{cl})}$ is compatible with the properties
\eqref{eq:AM13G}, since
\begin{equation}
[\hat E^{(\text{cl})},\hat L_k]=0 \,.
\label{eq:AM24}
\end{equation}
The detailed discussion of the Coulomb potential below will reveal that the condition \eqref{eq:AM23} is actually not necessary for realizing a periodic probability distribution. 
Periodic wave functions and their superpositions are a characteristic feature of quantum mechanics. 
It may therefore not be surprising that our quantum description of the probabilistic classical wave function will be an efficient tool for the understanding of its time evolution.

\section{Subsystems}
\label{sec:VIII}

\indent Rather generally, quantum systems can emerge as subsystems
of classical probabilistic systems \cite{CWPW, CWPW2}.
The probabilistic classical particle constitutes already a quantum system.
We can then look for (reduced) quantum systems as subsystems of this
quantum system, employing standard techniques from quantum
mechanics. For the free particle or the particle in a harmonic
potential we have already encountered such a subsystem. It is
quantum mechanics based on the operators $\hat x$ and $\hat p$.

\subsection*{Subsystems from closed sets of conserved observables}

\indent We can generalize this type of a quantum subsystem.
Consider a set of hermitian operators $\hat A_i$ which form a closed algebra
\begin{equation}
[\hat A_i,\hat A_j] = i f_{ijk}\hat A_k \,,
\label{eq:SU1}
\end{equation}
and commute with the Hamiltonian
\begin{equation}
[\hat A_i,H]=0 \,.
\label{eq:SU2}
\end{equation}
The expectation values of the observables $A_i$, which are represented by the operators $\hat A_i$, are conserved, $\partial_t \langle A_i\rangle = 0$.
These observables may be classical or statistical observables.

\indent The algebra \eqref{eq:SU1} generates a group of transformations acting on a complex wave function $\psi$. 
The infinitesimal transformations read
\begin{equation}
\delta \psi = i\alpha_i \hat A_i \psi \,,
\label{eq:SU3}
\end{equation}
with infinitesimal parameters $\alpha_i$. 
Consider now real representations of the transformation group.
In this case the transformation \eqref{eq:SU3} does not mix the real and imaginary components of $\psi$. 
We will work in this part mainly in the basis with a real wave function $q(z,v)$. 
In this basis the complex wave function $\psi$ is considered as a technical tool, with $q$ the real part of $\psi$.
For real representations, eq.~\eqref{eq:SU3} projects then to an infinitesimal transformation of the real wave function,
\begin{equation}
\delta q = i\alpha_i \hat A_i q \,.
\label{eq:SU4}
\end{equation}

\indent Different representations are characterized by the values of a set of Casimir operators $\hat C_s$ which can be constructed from $\hat A_i$. 
Since $[\hat C_s,H]=0$, a given sharp value of $\hat C_s$ does not change during the evolution.
An eigenstate of $\hat C_s$ remains an eigenstate during the evolution.
If $q$ belongs to a given representation at time $t_1$, it will belong to the same representation at any later time $t_2$. 
The representation may be reducible. 
A wave function with given values of the Casimir operators $\hat C_s$ can be a superposition of many wave functions with the same $\hat C_s$.
The ensemble of wave functions which are eigenfunctions to given eigenvalues $\hat C_s$ and therefore belonging to a given representation of a transformation group, form a quantum subsystem.
The Hamiltonian is block diagonal, with blocks acting separately on the different representations. 
If we take a given representation as a reduced quantum subsystem, its dynamics is determined by a reduced Hamiltonian, corresponding to the block for this subsystem.

\indent An example for such a reduced quantum system is given for a central potential by the eigenstates of $\hat L^2$ with $l=1$, which we have discussed above.
We can equivalently discuss this type of subsystem in the basis of the complex wave function $\varphi(z,s)$. 
For this purpose we consider the action of $H$ and the operators $\hat A_i$ on a general complex wave function $\psi(z,s)$ without imposing the constraint \eqref{eq:5}. 
Then the conjugate wave function $\bar\psi(z,s)=\psi^\ast(z,-s)$ also belongs again to a representation of the transformation group. 
The combinations
\begin{equation}
\varphi(z,s)
=
c_N\bigl(\psi(z,s)+\psi^*(z,-s)\bigr)\,,
\label{eq:SU5}
\end{equation}
with real normalization factor $c_N$, form again a reduced quantum subsystem. 
This allows $\psi$ to belong to a complex representation of the transformation group. 
The wave function $q(z,v)$ belongs to a real representation, which may be a sum of a complex representation and its conjugate representation. 
A finite sum of different representations forms still a quantum subsystem. 

\indent As a conceptually important example we may project $\psi$ on the subspace with positive or zero eigenvalues of $H$ by requiring
\begin{equation}
\text{sign}(H)\,\psi=\psi
\quad\text{or}\quad
H\psi=0 \,.
\label{eq:SU6}
\end{equation}
Then
\begin{equation}
\text{sign}(H)\,\bar\psi=-\bar\psi
\quad\text{or}\quad
H\bar\psi=0 \,.
\label{eq:SU7}
\end{equation}
The dynamics can be understood by a restriction to the reduced quantum subsystem defined by eq.~\eqref{eq:SU6}. 
It may often be difficult in practice to realize the condition \eqref{eq:SU6} since exact eigenvalues and eigenfunctions of $H$ are not known.
For the free particle or the harmonic potential the projection \eqref{eq:SU6} corresponds to the quantum particle, and the reduced Hamiltonian is given by $H^{(p)}$ in eq.~\eqref{eq:11}.

\subsection*{Statistical observables and Poisson brackets}

\indent The operators defining the algebra \eqref{eq:SU1} for a subsystem can represent classical or statistical observables. 
Among the statistical observables there is a set related to classical observables by Poisson brackets. 
They constitute a bridge to the historical "quantization" by the use of Poisson brackets.

\indent A classical observable $A^{(\text{cl})}$ is a function of the variables $z$ and $v$, represented by the operator
\begin{equation}
\hat A^{(\text{cl})}
=
A^{(\text{cl})}(z,v)\,\delta(z-z')\,\delta(v-v')\,.
\label{eq:SU8}
\end{equation}
One can associate to it a statistical quantum observable
$A^{(Q)}$, represented by the hermitian operator
\begin{equation}
\hat A^{(Q)}
=
i\left(
\frac{\partial A^{(\text{cl})}}{\partial z_k}\frac{\partial}{\partial v_k}
-
\frac{\partial A^{(\text{cl})}}{\partial v_k}\frac{\partial}{\partial z_k}
\right)
=
\bigl(\hat A^{(Q)}\bigr)^\dagger \,.
\label{eq:SU9}
\end{equation}
The action of $\hat A^{(Q)}$ on a function $B^{(\text{cl})}(z,v)$
yields the Poisson bracket of the functions $A^{(\text{cl})}$ and $B^{(\text{cl})}$,
\begin{align}
\hat A^{(Q)} B^{(\text{cl})}
&=
i\{A^{(\text{cl})},B^{(\text{cl})}\}
\nn \\
&=
i\left(
\frac{\partial A^{(\text{cl})}}{\partial z_k}\frac{\partial B^{(\text{cl})}}{\partial v_k}
-
\frac{\partial A^{(\text{cl})}}{\partial v_k}\frac{\partial B^{(\text{cl})}}{\partial z_k}
\right)\,.
\label{eq:SU10}
\end{align}
Correspondingly, the commutator of the operators
$\hat A^{(Q)}$ and $\hat B^{(Q)}$ obeys
\begin{equation}
[\hat A^{(Q)},\hat B^{(\text{cl})}]
=
i\,\{A^{(\text{cl})},B^{(\text{cl})}\} \,,
\label{eq:SU11}
\end{equation}
where the r.h.s.\ denotes the operator associated by eq.~\eqref{eq:SU8}
to the classical Poisson bracket.
The quantum observable associated to the product of two
classical observables is represented by the operator
\begin{equation}
\bigl(A^{(\text{cl})}B^{(\text{cl})}\bigr)^{Q}
=
A^{(\text{cl})}\hat B^{(Q)}
+
B^{(\text{cl})}\hat A^{(Q)} \,.
\label{eq:SU12}
\end{equation}
Finally, one finds for the commutator of two quantum observables
\begin{align}
\label{eq:SU13}
&[\hat A^{(Q)},\hat B^{(Q)}]
=
i\,\{A^{(\text{cl})},B^{(\text{cl})}\}^{(Q)} 
\\
&= -
\frac{\partial \{A^{(\text{cl})},B^{(\text{cl})}\}}{\partial z_k}
\frac{\partial}{\partial v_k}
+
\frac{\partial \{A^{(\text{cl})},B^{(\text{cl})}\}}{\partial v_k}
\frac{\partial}{\partial z_k}\,, \nn
\end{align}
where the r.h.s.\ is the quantum operator associated to the Poisson bracket by eq.~\eqref{eq:SU9}.
If the Poisson bracket vanishes for a pair of classical observables, the associated quantum observables are represented by commuting operators.

\indent The Hamiltonian $H$ is the quantum operator associated to the classical energy
\begin{equation}
H = \bigl(E^{(\text{cl})}\bigr)^Q \,.
\label{eq:SU14}
\end{equation}
The quantum angular momentum is associated to the classical angular momentum
\begin{equation}
\hat L_k = \bigl(L_k^{(\text{cl})}\bigr)^Q \,.
\label{eq:SU15}
\end{equation}
The commutator relations for $\hat L_k$ and $H$ can be
understood from the relations above. Possible subsystems based
on the algebra \eqref{eq:SU1} may be found by simple relations
for Poisson brackets for suitable classical observables.

\section{Probabilistic classical particle in the Coulomb potential}
\label{sec:IX}

\indent In this section we apply the quantum formalism to the probabilistic
description of a classical particle in a potential $V\sim 1/r$, see for example ref \cite{SUBA}.
This covers a wide range of scales. 
On macroscopic scales we describe dust around a star in the limit where the 
interaction among dust particles and radiation effects can be neglected, 
and only the gravitational field of the star governs the dynamics of the particles.
On the microscopic scale we discuss the dynamics of a classical
electron (without spin) in the Coulomb potential of a point charge. 
This differs from the standard quantum system for a quantum electron in the hydrogen atom.
For both the quantum electron and the "classical electron" the same quantum formalism is used. 
The difference is no longer conceptual -- it only resides from a different Hamiltonian. 
This should help to gain a clearer view on the peculiarities of the quantum electron, as well as on the common features to both settings.

\subsection*{Scales for the Coulomb potential}

\indent We consider the Coulomb potential in units with $\hbar = m = 1$,
\begin{equation}
V(r) = - \frac{\kappa}{r}\,,
\quad
F_k = - \frac{\kappa z_k}{r^3} \,.
\label{eq:C1}
\end{equation}
It obtains by the rescaling \eqref{eq:1A} from the potential in dimensionful units.
For the Kepler problem we describe dust or planets around a star
with mass $M_s$, and $G$ the gravitational constant,
\begin{align}
&V_{\mathrm{ph}} = mV = - \frac{GM_s m}{r_{\text{ph}}}
= - \frac{\kappa_{\mathrm{g}}}{r_{\text{ph}}}\,,
\nn \\
&\kappa_{\mathrm{g}} = GM_s m = \frac{c}{8\pi} \frac{M_s m}{M^2}  \,.
\label{eq:C1A}
\end{align}
with $M$ the reduced Planck mass and $c$ the light velocity.
For the classical electron we employ
\begin{equation}
V_{\text{ph}} = - \frac{\alpha c}{r_{\text{ph}}}=-\frac{\kappa_{\mathrm{e}} }{r_{\text{ph}}}\,,
\quad
\kappa_{\mathrm{e}} = \alpha c \,.
\label{eq:C1B}
\end{equation}
with $\alpha$ the dimensionaless fine structure constant.
In our units $c$ is not dimensionless and $\kappa$ has units $t^{-1/2}$, as a velocity or inverse distance. 
For dust particles the ratio $\kappa_{\mathrm{g}}/\kappa_{\mathrm{e}}$ is huge. 
The quantity $\kappa$ can be set to one by a further rescaling of time and $v$,
\begin{equation}
\bar t = \kappa^2 t\,,
\quad
\bar z = \kappa z\,,
\quad
\bar v = \frac{v}{\kappa}\,,
\quad
\bar E = \frac{E}{\kappa^2}\,,
\quad
\bar H = \frac{H}{\kappa^2}\,.
\label{eq:C9}
\end{equation}
This shows that the dynamics is the same for macroscopic dust and the
classical electron once we use the rescaled variables $\bar z$, $\bar v$ and $\bar t$.
We will keep here the constant $\kappa$ for purposes of bookkeeping.

\indent With these units the classical energy and the Liouville operator read ($v^2 = v_k v_k$)
\begin{equation}
E^{(\text{cl})} = \frac{v^2}{2} - \frac{\kappa}{r}\,,
\quad
\mathcal{L}
=
v_k \frac{\partial}{\partial z_k}
- \frac{\kappa z_k}{r^3}\frac{\partial}{\partial v_k} \,.
\label{eq:C2}
\end{equation}
We describe the dynamics here in terms of a complex wave function $\psi$
\begin{equation}
i\partial_t \psi = H\psi = - i \mathcal{L}\psi \,,
\label{eq:C2A}
\end{equation}
and extract $q$ at the end as the real part of $\psi$.
(The evolution equation does not mix real and imaginary parts of $\psi$.)
Besides the usual rotation invariance the dynamics is
invariant under scale transformations,
\begin{equation}
t \to \alpha t\,,
\quad
z \to \alpha^{2/3} z\,,
\quad
v \to \alpha^{-1/3} v \,.
\label{eq:C2B}
\end{equation}
They transform the classical and quantum energy according to
\begin{equation}
E^{(\text{cl})} \to \alpha^{-2/3} E^{(\text{cl})}\,,
\quad
H \to \alpha^{-1} H \,.
\label{eq:C2C}
\end{equation}

\indent The trajectories of classical point particles are closed orbits
with frequencies $\omega^{(\text{cl})}$ related to the classical energy $E<0$,
\begin{equation}
\omega^{(\text{cl})} = \kappa^{-1}(-2E)^{3/2} \,.
\label{eq:C2D}
\end{equation}
Here the energy dependence is dictated by scale symmetry \eqref{eq:C2C},
$\omega^{(\text{cl})} \to \alpha^{-1}\omega^{(\text{cl})}$, and the power of
$\kappa$ follows from dimensions, with $\omega^{(\text{cl})}\sim t^{-1}$,
$E^{(\text{cl})}\sim t^{-1}$. Periodic wave functions or probability
distributions can occur for a fixed sharp classical energy $E^{(\text{cl})}=\bar E$. 
For a possible subsystem with fixed classical energy the wave functions are given by
\begin{equation}
\psi = \tilde\psi\, \delta^{\frac{1}{2}} \bigl(E^{(\text{cl})}-\bar E\bigr) \,.
\label{eq:C2E}
\end{equation}
The reduced wave function $\tilde\psi$ depends on five variables which parametrize in phase space the shells with fixed classical energy. 
The classical energy is conserved.
The time evolution of $\psi(t)$ is determined for this subsystem by the simultaneous solution of
\begin{equation}
i\partial_t \psi = H\psi\,,
\quad
\hat E^{(\text{cl})} \psi = \bar E \psi \,.
\label{eq:C2F}
\end{equation}
Simultaneous eigenstates of $H$ and $\hat E^{(\text{cl})}$ are possible since $[H,\hat E^{(\text{cl})}]=0$. 

\subsection*{Conserved observables}

\indent We will focus on subsystems based on conserved quantities. 
Conserved quantities are observables, including statistical observables, for which the associated operators commute with $H$.
We first have the classical conserved quantities, namely angular momentum $\vec{L}^{(\text{cl})}$ and the Runge--Lenz vector $\vec{A}^{(\text{cl})}$. 
In the basis $\psi(z,v)$ these are simply functions of the phase-space coordinates $z$, $v$, where we omit the $\delta$-functions in eq.~\eqref{eq:SU8},
\begin{align}
\label{eq:C3}
L_k^{(\text{cl})} &= \varepsilon_{klm} z_l v_m \,, \\
A_k^{(\text{cl})}
&=
\varepsilon_{klm} v_l L_m^{(\text{cl})} - \frac{\kappa z_k}{r}
=
v^2 z_k - (vz) v_k - \frac{\kappa z_k}{r} \,. \nn
\end{align}
They obey
\begin{align}
L^{(\text{cl})2}
&=
L_k^{(\text{cl})} L_k^{(\text{cl})}
=
r^2 v^2 - (zv)^2 \,, \nonumber\\
L_k^{(\text{cl})} A_k^{(\text{cl})} &= 0 \,, \nonumber\\
A^{(\text{cl})2}
&=
A_k^{(\text{cl})} A_k^{(\text{cl})}
=
\kappa^2 + 2E^{(\text{cl})} L^{(\text{cl})2} \,.
\label{eq:C4}
\end{align}
We can therefore express the classical energy $E^{(\text{cl})}$ in terms of
$A^{(\text{cl})2}$ and $L^{(\text{cl})2}$.

\indent Associating to $z_k$ and $v_k$ the operators $\hat z_k$ and $\hat v_k$
the operators for the classical angular momentum and the classical
Runge--Lenz vector commute with $H$
\begin{equation}
[H,\hat L_k^{(\text{cl})}] = 0\,,
\quad
[H,\hat A_k^{(\text{cl})}] = 0 \,.
\label{eq:C5}
\end{equation}
Both $L_k^{(\text{cl})}$ and $A_k^{(\text{cl})}$ are conserved quantities.
Any wave function $q(L_k,A_k)$ which only depends on $\vec L^{(\text{cl})}$
and $\vec A^{(\text{cl})}$ is static.
Once we have associated operators to observables we can make arbitrary basis transformations.

\indent Besides the classical conserved quantities there are also conserved statistical observables.
For a rotation symmetric potential the quantum angular momentum is conserved. 
Similar to the quantum angular momentum we define the quantum
Runge--Lenz vector as a statistical observable. 
Its associated operator is given according to eq.~\eqref{eq:SU9} by
\begin{align}
\hat A_k &= i\left( \frac{\partial A_k^{(\text{cl})}}{\partial z_m}\frac{\partial}{\partial v_m}
- \frac{\partial A_k^{(\text{cl})}}{\partial v_m} \frac{\partial}{\partial z_m}   \right)
\nn \\
&=- i \Biggl[
\bigl(2 z_k v_m - v_k z_m \bigr)\frac{\partial}{\partial z_m} -(vk)\frac{\partial}{\partial z_k}
- v^2 \frac{\partial}{\partial v_k}
\nn \\
&\hspace{1cm}+ v_k v_m \frac{\partial}{\partial v_m}
+ \frac{\kappa}{r}\frac{\partial}{\partial v_k}
- \frac{\kappa}{r^3} z_k z_m \frac{\partial}{\partial v_m}
\Biggr] \,.
\label{eq:C6}
\end{align}
Being a vector under rotations, $\hat A_k$ obeys the usual commutator relation
\begin{equation}
[\hat L_k,\hat A_l] = i\varepsilon_{klm}\hat A_m \,.
\label{eq:C7}
\end{equation}
The commutator with the Hamiltonian vanishes
\begin{equation}
[H,\hat A_k] = 0 \,.
\label{eq:C8}
\end{equation}
such that the quantum Runge--Lenz vector is a conserved quantity. 

\indent One also finds the commutator
\begin{align}
[\hat A_k,\hat A_l]
&=-2i\,\varepsilon_{klm}
\left(
E^{(\text{cl})} L_m^{(\text{cl})}
\right)^{(Q)}
\nn \\
&=-2i\,\varepsilon_{klm} \left( E^{(\text{cl})} \hat L_m  + L_m^{(\text{cl})}H\right)\,.
\label{eq:C8A}
\end{align}
Eqs.~\eqref{eq:C8} and \eqref{eq:C8A} follow from the Poisson brackets
\begin{equation}
\{A_k^{(\text{cl})},E^{(\text{cl})}\} = 0\,,
\quad
\{A_k^{(\text{cl})},A_l^{(\text{cl})}\}
= -2E^{(\text{cl})} \varepsilon_{klm} L_m^{(\text{cl})} \,.
\label{eq:C13A}
\end{equation}
From eq.~\eqref{eq:C8A} one infers the relation
\begin{equation}
(L^2)^{(\text{cl})} H
=
\frac{i}{4}\varepsilon_{klm} L_m^{(\text{cl})}[\hat A_k,\hat A_l]
-
E^{(\text{cl})} L_k^{(\text{cl})} \hat L_k \,,
\label{eq:C14}
\end{equation}
with
\begin{equation}
(L^2)^{(\text{cl})} = r^2 v^2 - (zv)^2,
\quad
L_k^{(\text{cl})} \hat L_k
=
\frac{1}{2}\bigl((L^2)^{(\text{cl})}\bigr)^{(Q)} \,.
\label{eq:C15}
\end{equation}
Using eq.~\eqref{eq:SU10} together with $\{L^2,E\}=0$ one also finds
\begin{equation}
(L^2)^{(\text{cl})} H + L_k^{(\text{cl})} \hat L_k \hat E^{(\text{cl})}
=
\frac{i}{4}\varepsilon_{klm} L_m^{(\text{cl})}[\hat A_k,\hat A_l] \,.
\label{eq:C16}
\end{equation}

\indent A closed set of operators forms an algebra with an associated group
of transformations of the wave function. We are interested in such
algebras for operators commuting with $H$ and $\hat E^{(\text{cl})}$.
In this case one can find wave functions which form representations of the transformation group.
The operators for classical observables all mutually commute and trivially form an algebra.
One can find simultaneous eigenstates of $\hat L_k^{(\text{cl})}$,
$\hat A_k^{(\text{cl})}$, $\hat E^{(\text{cl})}$ and $H$. 
They correspond to wave functions with support only on a given orbit of
a point particle fixed by the values of the classical angular momentum and Runge--Lenz vector. 
These solutions are relevant for the motion of planets in astrophysics, but not for microphysical settings.

\indent The quantum angular momentum forms an algebra which is more adapted
to microphysics. We can therefore find wave functions which belong
to a given representation of the rotation group. They are characterized
by a fixed value of $\hat L_k \hat L_k$. The sharp value of this
statistical observable does not require a fixed value of the squared
classical angular momentum. With eqs.~\eqref{eq:AM8}, \eqref{eq:AM9} 
wave functions belonging to a given representation of the rotation
group and with fixed classical squared angular momentum
$(L^{(\text{cl})})^2$ are possible, but a fixed $(L^{(\text{cl})})^2$ 
is not necessary for an eigenstate of $\hat L_k \hat L_k$.

\subsection*{$SO(4)$-symmetry}

\indent For the subspace of wave functions with negative classical energy,
\begin{equation}
\text{sign}\bigl(\hat E^{(\text{cl})}\bigr) \psi = - \psi \,,
\label{eq:C18}
\end{equation}
we can extend the algebra by including quantum operators for classical functions
which are proportional to the Runge--Lenz vector.
We define classical observables
\begin{equation}
B_k = \frac{1}{\sqrt{-2E}}\,A_k \,,
\label{eq:C19}
\end{equation}
where we take the positive root. 
With Poisson brackets
\begin{equation}
\{f(E),A_k\} = 0\,,
\quad
\{f(E),L_k\} = 0 \,,
\label{eq:C20}
\end{equation}
one obtains the Poisson brackets for $B_k$,
\begin{equation}
\{L_k,B_l\} = \varepsilon_{klm} B_m\,,
\quad
\{B_k,B_l\} = \varepsilon_{klm} L_m \,.
\label{eq:C21}
\end{equation}

\indent The statistical observables associated to $B_k$ correspond to the operators
\begin{equation}
\hat B_k
=
(-2E^{(\text{cl})})^{-\frac{1}{2}}\hat A_k
+
(-2E^{(\text{cl})})^{-\frac{3}{2}} A_k^{(\text{cl})} H \,.
\label{eq:C21A}
\end{equation}
They obey the commutation relations
\begin{align}
[\hat L_k,\hat B_l] &= i\varepsilon_{klm}\hat B_m \,, \quad
[\hat B_k,\hat B_l] = i\varepsilon_{klm}\hat L_m \,, \nonumber\\
[\hat B_k,H] &= 0 \,, \qquad\qquad
[\hat B_k,\hat E^{(\text{cl})}] = 0 \,.
\label{eq:C22}
\end{align}
The generators $\hat L_k$, $\hat B_k$ therefore form a closed algebra.
The associated transformation group is $SO(4)$. 
This can be seen by defining
\begin{equation}
\hat J_k^{+} = \frac{1}{2}\bigl(\hat L_k + \hat B_k\bigr)\,,
\quad
\hat J_k^{-} = \frac{1}{2}\bigl(\hat L_k - \hat B_k\bigr) \,.
\label{eq:C23}
\end{equation}
The commutator relations
\begin{align}
[\hat J_k^{+},\hat J_l^{+}] &= i\varepsilon_{klm}\hat J_m^{+} \,, \quad
[\hat J_k^{-},\hat J_l^{-}] = i\varepsilon_{klm}\hat J_m^{-} \,, \nonumber\\
[\hat J_k^{+},\hat J_l^{-}] &= 0 \,, \qquad\qquad
[\hat L_k,\hat J_l^{\pm}] = i\varepsilon_{klm}\hat J_m^{\pm} \,,
\label{eq:C24}
\end{align}
reflect the structure $SO(4) \hat = SO(3)\times SO(3)$.
This is the characteristic symmetry group for the quantum mechanics of the hydrogen atom. 

\indent The representations are eigenstates of the Casimir operators
\begin{equation}
\hat J^{+2} = \hat J_k^{+}\hat J_k^{+}\,,
\quad
\hat J^{-2} = \hat J_k^{-}\hat J_k^{-} \,.
\label{eq:C25}
\end{equation}
From the commutation relations \eqref{eq:C24} we know that the
eigenvalues of $\hat J^{\pm2}$ are $j_{\pm}(j_{\pm}+1)$, with integer or half integer $j_{\pm}$.
In terms of $\hat L_k$ and $\hat B_k$ these Casimir operators read
\begin{equation}
\hat J^{\pm2}
=
\frac{1}{4}
\left(
\hat L^2 + \hat B^2
\pm
\bigl(\hat L_k \hat B_k + \hat B_k \hat L_k\bigr)
\right) \,.
\label{eq:C26}
\end{equation}
All quantities in eq.~\eqref{eq:C26} are rotation invariant,
\begin{equation}
[\hat L_k,\hat B^2] = 0\,,
\quad
[\hat L_k,\hat J^{\pm2}] = 0 \,.
\label{eq:C27}
\end{equation}
On the other hand, the non-vanishing commutators
\begin{align}
[\hat B_k,\hat L^2]
&=
i\varepsilon_{klm}\bigl(\hat L_l \hat B_m + \hat B_m \hat L_l\bigr) \,, \nonumber\\
[\hat J_k^{\pm},\hat L^2]
&=
i\varepsilon_{klm}\bigl(\hat L_l \hat J_m^{\pm} + \hat J_m^{\pm}\hat L_l\bigr) \,,
\label{eq:C28}
\end{align}
indicate that $SO(4)$ transformations can change the representation of the rotation subgroup.

\indent The $SO(4)$ representations form quantum subsystems which are closed under the evolution. 
The quantum observables formed from $\hat J_k^{\pm}$ are conserved in time. 
With $[\hat J_k^{\pm},\hat E^{(\text{cl})}]=0$ the $SO(4)$-transformations act
within shells of fixed classical energy in phase space. 
We can therefore consider the subsystem of a given $SO(4)$-representation with fixed classical energy \eqref{eq:C2}, \eqref{eq:C2F}.
We are particularly interested in the eigenstates of the Hamiltonian
\begin{equation}
H\psi = \omega \psi \,.
\label{eq:C29}
\end{equation}
The $SO(4)$ transformations act in the space of eigenstates with given $\omega$. 
These eigenstates belong to $SO(4)$-representations. 
This produces the well known degeneracies for the hydrogen atom. 
For example, the rotation singlet $2s$ and the triplet $2p$ have the same
$\omega$, forming four degenerate states. 
(We do not have here the additional two-fold degeneracy due to the electron spin in the hydrogen atom.)

\subsection*{Discrete symmetry}

\indent We may also investigate the action of discrete transformations. 
The parity transformation acts as
\begin{align}
P:\qquad
& z_k \to - z_k\,,\quad
v_k \to - v_k\,,\quad
E^{(\text{cl})} \to E^{(\text{cl})}\,, \nn\\
& H \to H\,, \quad 
L_k^{(\text{cl})} \to L_k^{(\text{cl})}\,,\quad
B_k^{(\text{cl})} \to - B_k^{(\text{cl})}\,,\nn \\
&\hat L_k \to \hat L_k\,,\quad
\hat B_k \to - \hat B_k\,,\quad
\hat J_k^{+} \leftrightarrow \hat J_k^{-} \,.
\label{eq:C30}
\end{align}
A given $SO(4)$-representation is an eigenstate of $\hat J^{+2}$ and $\hat J^{-2}$. 
We demand that the parity transformation does not change the representation. 
From
\begin{align}
\hat J^{+2} P\psi
&=
P\hat J^{-2}\psi
=
P\,j^{-}(j^{-}+1)\psi \nn \\
&=
j^{-}(j^{-}+1) P\psi
=
j^{+}(j^{+}+1) P\psi 
\label{eq:C31}
\end{align}
we conclude $j^{+}=j^{-}$ or
\begin{equation}
\hat J^{+2}\psi = \hat J^{-2}\psi \,.
\label{eq:C32}
\end{equation}
We therefore deal with the $(n,n)$-representation of $SO(3)\times SO(3)$. 
These are real representations which can act on the real wave function $q$. 
For example, the representation $(2,2)$ is composed of a triplet and a singlet of the rotation group.

\indent We may also consider the transformation $\tilde T$ which is related to time reversal,
\begin{align}
\tilde T:\qquad
& z_k \to z_k\,,\quad
v_k \to - v_k\,,\quad
E^{(\text{cl})} \to E^{(\text{cl})}\,,\quad \nn\\
& H \to - H\,, \quad
L_k^{(\text{cl})} \to - L_k^{(\text{cl})}\,,\quad
B_k^{(\text{cl})} \to B_k^{(\text{cl})}\,,\nn \\
&\hat L_k \to \hat L_k\,,\quad
\hat B_k \to - \hat B_k\,,\quad
\hat J_k^{+} \leftrightarrow \hat J_k^{-} \,.
\label{eq:C33}
\end{align}
This restricts the possible relation between $H$ and other operators, as for eq.~\eqref{eq:C16}. 
The classical and quantum Runge--Lenz vectors $A_k^{(\text{cl})}$, $\hat A_k$ 
transform as $B_k^{(\text{cl})}$ and $\hat B_k$ under $P$ and $\tilde T$.

\indent The values of $j_{+}=j_{-}$ can be half-integer and we take
\begin{equation}
j_{+} = j_{-} = \frac{n-1}{2} 
\label{eq:C34}
\end{equation}
with integer $n$. 
For $n=1$ we deal with a singlet of $SO(4)$, and for $n=2$ with the four-component vector $(2,2)$. 
For $n=3$ one has the representation $(3,3)=1+3+5$, 
where we have decomposed into representations of the rotation group $SO(3)$. 
The degeneracies of the quantum-energy eigenstates for the quantum electron in the hydrogen
atom appear naturally for the probabilistic classical electron in the Coulomb potential.

\subsection*{Scale invariant variables}

\indent We next introduce scale-invariant and dimensionless variables
\begin{equation}
\sigma_k = \frac{1}{\kappa \varepsilon^2} z_k\,,
\quad
\zeta_k = \varepsilon v_k\,,
\quad
\varepsilon = \frac{1}{\sqrt{-2E}} \,.
\label{eq:S1}
\end{equation}
The variable $\sigma_k$ measures the position in the units of the
Kepler length scale $(\kappa\varepsilon^2)$ (semi-major axis scale in the Kepler problem),  and the velocity in terms of the natural energy scale in the Kepler problem.
With
\begin{equation}
\rho^2 = \sigma_k \sigma_k\,,
\quad
\zeta^2 = \zeta_k \zeta_k\,,
\quad
\tilde q = \sigma_k \zeta_k \,,
\label{eq:S3}
\end{equation}
one has the relation
\begin{equation}
    \label{eq:S1A}
    \frac{1}{\rho} =  \frac{1+\zeta^2}{2}\,.
\end{equation}
This decomposes phase space into $\varepsilon$ and five independent scale invariant deminsionless variables (angles).

\indent We can use the invariant building blocks
\begin{align}
&\mathcal{H}=\kappa \varepsilon^3 H
=
- i\left(
\zeta_k \frac{\partial}{\partial \sigma_k}
-
\frac{\sigma_k}{\rho^3}\frac{\partial}{\partial \zeta_k}
\right) \,, \nonumber\\
&\frac{1}{\kappa} A_k^{(\text{cl})}
=
\left(
\zeta^2 - \frac{1}{\rho}
\right)\sigma_k
-
\tilde q\, \zeta_k \,.
\label{eq:S2}
\end{align}
Here we employ
\begin{align}
\frac{\partial}{\partial z_k}
&=
\frac{1}{\kappa\varepsilon^2}
\left\{
\frac{\partial}{\partial \sigma_k}
+
\frac{\sigma_k}{\rho^3}
\left(
\zeta_m \frac{\partial}{\partial \zeta_m}
-
2\sigma_m \frac{\partial}{\partial \sigma_m}
\right)
\right\} \,, \nonumber\\
\frac{\partial}{\partial v_k}
&=
\varepsilon
\left\{
\frac{\partial}{\partial \zeta_k}
+
\zeta_k
\left(
\zeta_m \frac{\partial}{\partial \zeta_m}
-
2\sigma_m \frac{\partial}{\partial \sigma_m}
\right)
\right\} \,.
\label{eq:S4}
\end{align}
Similarly, one finds
\begin{align}
\varepsilon \hat A_k
=i \Biggl\{
&\left( \zeta^2 - \frac{1}{\rho}\right)\frac{\partial}{\partial \zeta_k}
+\tilde q \frac{\partial}{\partial \sigma_k} \nn \\
&+\zeta_k
\left( \sigma_m \frac{\partial}{\partial \sigma_m}
-\zeta_m \frac{\partial}{\partial \zeta_m} \right) \nn \\
&+\sigma_k
\left( \frac{1}{\rho^3}\sigma_m \frac{\partial}{\partial \zeta_m}
-2\zeta_m \frac{\partial}{\partial \sigma_m}\right)
\Biggr\} \,.
\label{eq:S5}
\end{align}

\indent This expresses the $SO(4)$-generator $\hat B_k$ in terms of $\sigma$ and $\zeta$ and their derivatives without $\varepsilon$ or $\kappa$ appearing anymore
\begin{align}
\label{eq:S6}
&\hat B_k
=
\varepsilon \hat A_k
+
\varepsilon^3 A_k^{(\text{cl})} H \\
&=
i\Biggl\{
\frac{1-\rho}{\rho}\frac{\partial}{\partial \zeta_k}
+
\tilde q \frac{\partial}{\partial \sigma_k}
+
\zeta_k
\left(
\sigma_m \frac{\partial}{\partial \sigma_m}
-
\zeta_m \frac{\partial}{\partial \zeta_m}
\right) \nn \\
&+
\left(
\frac{\sigma_k}{\rho}
-
\tilde q\,\zeta_k
\right)
\frac{1}{\rho^3}\,
\sigma_m \frac{\partial}{\partial \zeta_m}
-
\left(
\frac{1+\rho}{\rho}\sigma_k
-
\tilde q\,\zeta_k
\right)
\zeta_m \frac{\partial}{\partial \sigma_m}
\Biggr\}\,. \nn
\end{align}
Also $\hat L_k$ can be expressed in terms of $\zeta_k$ and $\sigma_k$, without dependence on $\varepsilon$ and $\kappa$.

\section{Hydrogen atom from probabilistic classical particle}
\label{sec:X}

\indent The hydrogen atom for a spinless electron can be obtained as a reduced quantum system.
It is a subsystem within the probabilistic description for a classical particle in the Coulomb (or Newton) potential.
More precisely, the restricted set of those observables which are compatible with this subsystem shares the properties of the quantum electron in a central potential except for the spin.

\subsection*{Constrained subsystem}

\indent Let us define two statistical observables represented by the operators
\begin{equation}
\hat D_Q = \hat{B}_k \,\hat{L}_k\,,
\label{eq:Q1}
\end{equation}
and
\begin{equation}
\hat C_Q = - \frac{2 H}{\kappa^2}\left(\hat L^2+\hat B^2+1\right)\,.
\label{eq:Q2}
\end{equation}
Both are conserved observables,
\begin{equation}
[H,\hat D_Q]=0\,,
\quad
[H,\hat C_Q]=0\,.
\label{eq:Q3}
\end{equation}
They also commute among each other,
\begin{equation}
[\hat C_Q,\hat D_Q]=0\,.
\label{eq:Q4}
\end{equation}

\indent The reduced quantum subsystem is defined by imposing two constraints
\begin{equation}
\hat D_Q \psi = 0\,,
\label{eq:Q6}
\end{equation}
and
\begin{equation}
\hat C_Q \psi = \psi\,.
\label{eq:Q7}
\end{equation}
In other words, the subsystem consists of simultaneous eigenfunctions of $\hat D_Q$ and $\hat C_Q$, with eigenvalues zero and one. 
Due to eq.~\eqref{eq:Q3} the subsystem is closed under time evolution. 
The constraint \eqref{eq:Q6} restricts the wave function to be composed of $(n,n)$-representations of the $SO(4)$-transformation group since it implies
\begin{equation}
\left(\hat J_+^2-\hat J_-^2\right)\psi = 0\,.
\label{eq:Q8}
\end{equation}
As mentioned before, these are precisely the representations appearing in the quantum hydrogen atom. 
The second condition \eqref{eq:Q7} may be named the ``quantum constraint'', even though it does not introduce the quantum structure but rather reduces a larger quantum system to a reduced quantum subsystem.
The spectrum of possible eigenvalues of $\hat C_Q$ is continuous, covering both positive and negative real numbers. 
This follows from the continuous spectrum of $H$, and the presence of pairs of negative and positive eigenvalues of $H$. 
With $\hat L^2+\hat B^2+1$ a strictly positive operator, the constraint \eqref{eq:Q7} enforces negative eigenvalues of $H$.
We have chosen the power $\kappa^{-2}$ such that $\hat C_Q$ is dimensionless. 
On the other hand, $\hat C_Q$ is not a scale-invariant operator, since $H/\kappa^2$ is not scale invariant. 
The constraint \eqref{eq:Q7} breaks scale invariance, and the reduced subsystem will no longer be scale invariant.

\indent The constraint \eqref{eq:Q7} has an important consequence for the allowed eigenvalues of $H$ which can occur for the reduced quantum system. 
The possible eigenvalues of $\hat L^2+\hat B^2+1$ are given by $n^2$ with integer $n$. 
This follows from the condition \eqref{eq:Q6} with
\begin{equation}
\hat L^2+\hat B^2 = 4\hat J^{+2} = 4\hat J^{-2} = 4 j_\pm(j_\pm+1)\,,
\label{eq:Q9}
\end{equation}
and half integer or integer $j_\pm$ labelled by
\begin{equation}
j_\pm = \frac{1}{2}(n-1)\,,
\quad
4j_\pm(j_\pm+1)=n^2-1\,.
\label{eq:Q10}
\end{equation}
For eigenvalues of $\hat J^{\pm 2}$ the possible frequencies $\omega$ are restricted by the quantum constraint \eqref{eq:Q7}, with
\begin{equation}
H\psi=\omega\,\psi\,,
\quad
\hat J^{\pm 2}\psi = j_\pm(j_\pm+1)\psi\,,
\label{eq:Q11}
\end{equation}
implying
\begin{equation}
    \label{eq:Q11A}
    -\frac{2\omega n^2}{\kappa^2}=1\,.
\end{equation}
This results in
\begin{equation}
\omega= - \frac{\kappa^2}{2n^2}\,.
\label{eq:Q12}
\end{equation}
We may restore dimensions by multiplication with the appropriate power of the mass $m$, using eq.~\eqref{eq:C1B} with $m$ the electron mass.
This yields the well known quantum energy levels of the hydrogen atom
\begin{equation}
E_n = m\omega = -\frac{1}{2}\frac{\alpha^2 m c^2}{n^2}\,.
\label{eq:Q13}
\end{equation}

\indent The quantum conditions \eqref{eq:Q6}, \eqref{eq:Q7} are invariant under $SO(4)$ rotations, since
\begin{equation}
[\hat L_m,\hat B_k \, \hat L_k]=0\,,
\quad
[\hat B_m,\hat B_k \, \hat L_k]=0\,.
\label{eq:Q14}
\end{equation}
The eigenstates of $H$ are $SO(4)$-representations in the $(n,n)$-representation. 
The degeneracy is $n^2$, as for the (spinless) electron in the hydrogen atom. 
The rotation group is the diagonal subgroup of $SO(4)=SO(3)\times SO(3)$, with generators
\begin{equation}
\hat L_k = \hat J_{+k} + \hat J_{-k}\,.
\label{eq:Q15}
\end{equation}
This yields the same values for $\hat L^2=l(l+1)$ with integer $l$ as for the quantum electron.

\subsection*{Classical energy}

\indent The operator $\hat{E}^{(\text{cl})}$ for the classical energy commutes with $H$, $\hat{D}_Q$ and $\hat{C}_Q$.
One could construct an "extended hydrogen subsystem" for which $\hat{E}^{(\text{cl})}$ is an additional observable.
The quantum hydrogen atom does not have $\hat{E}^{(\text{cl})}$ as an observable.
One may therefore construct a subsystem which no longer contains information about the classical energy.
Since $\hat{E}^{(\text{cl})}$ commutes with $H$ the classical energy is conserved. 
In principle, one may therefore have sharp values of $E^{(\text{cl})}$ and $H$ simultaneously. 
On the other hand, the constraints \eqref{eq:Q6}, \eqref{eq:Q7} do not involve $E^{(\text{cl})}$.
There is therefore no need that $E^{(\text{cl})}$ assumes a sharp value.
We will argue that only a restricted set of values for $E^{(cl)}$ is compatible with an eigenfunction of $H$.
While no uncertainty relation based on a non-vanishing commutator forbids combinations of arbitrary eigenvalues $(\omega, \bar E)$ of $H$
and $\hat E^{(\text{cl})}$, this does not imply that eigenfunctions exist for arbitrary combinations.

\indent The evolution in phase space does not mix different energy shells characterized by different values of $E^{(\text{cl})}=\bar E$. 
For an eigenstate of $H$ with eigenvalue $\omega$ the contribution from each energy shell has to be periodic in time with period $T=2\pi/\omega$ and should involve only the frequency $\omega$. 
For the motion within a given energy shell the only possible frequencies are
\begin{equation}
\omega_k(\bar E)
=
k\,\omega^{(\text{cl})}(\bar E)
=
\frac{k}{\kappa}(-2\bar E)^{3/2}\,,
\label{eq:Q19}
\end{equation}
with positive integer $k$.
Therefore a given energy shell can contribute to an eigenstate of $H$ with eigenvalue $\omega_n$ only if $\omega_k (\bar E) = \omega_n$.
This restricts the possible values of $\bar E(\omega)$ which contribute for a given $\omega$ to a discrete set $\bar E_k (\omega)$,
\begin{equation}
\bar E_k(\omega)
= -\frac{1}{2}\left|\frac{\kappa\omega_n}{k}\right|^{2/3}\,.
\label{eq:Q20}
\end{equation}

\indent Correspondingly, for a given $n$ one has only contributions from classical energies
\begin{equation}
\bar E_k(n) = -\kappa^2 (32 n^4)^{-1/3} k^{-2/3}\,,
\label{eq:Q21}
\end{equation}
or
\begin{equation}
\epsilon_k(n)
=\frac{1}{\sqrt{-2\bar E_k(n)}}
=\frac{1}{\kappa}\bigl(2n^2 k\bigr)^{1/3}\,.
\label{eq:Q22}
\end{equation}
An infinite set of classical energies contributes to a given eigenvalue of $H$.
There is no one-to-one correspondence between oscillation frequencies and values of the classical energy.
For $k \to \infty$ the classical energy $\bar{E}_k(n)$ approaches zero.

\subsection*{Wave function for subsystems}

\indent An arbitrary complex wave function of the reduced quantum system can be written as a superposition of eigenfunctions of $H$, $\hat L^2$ and $\hat L_3$, as well known from the quantum electron,
\begin{align}
&\psi(t) = \sum_{n l m} c_{nlm} e^{-i\omega_n t}\psi_{nlm}\,,
\quad
H\psi_{nlm}=-\frac{\kappa^2}{2n^2}\psi_{nlm}\,,
\nn \\
&\hat L^2\psi_{nlm}=l(l+1)\psi_{nlm}\,,
\quad
\hat L_3\psi_{nlm}=m\psi_{nlm}\,.
\label{eq:Q16}
\end{align}
The wave function of the reduced quantum system belongs to a similar Hilbert space as the one for the (spinless) quantum electron. 
The real wave function is given by taking the real part of $\psi$,
\begin{align}
q(t)
=
\sum_{nlm}
\Big[
&\Re\bigl(c_{nlm}\psi_{nlm}\bigr)\cos(\omega_n t)
\nn \\
&-\Im\bigl(c_{nlm}\psi_{nlm}\bigr)\sin(\omega_n t)
\Big]\,.
\label{eq:Q17}
\end{align}
The corresponding probability distribution $w(t)=q^2(t)$ describes oscillating distributions in phase space.

\indent One would like to find expressions of $\psi_{nlm}$ as functions of $z$ and $v$.
This would allow for an explicit construction of probability distributions which realize the quantum states of the hydrogen atom.
These wave functions may not be unique -- different classical wave functions may be mapped to the same $\psi_{nlm}$.
Phase space is six-dimensional, parameterized by $(z,v)$ or, equivalently, by five independent scale invariant variables and $\varepsilon$.
The two constraints \eqref{eq:Q6}, \eqref{eq:Q7} leave a four-dimensional subspace.
We may therefore start with the "extended hydrogen subsystem" for which $\varepsilon$ is still a variable.

\indent A general wave function for this quantum system can be written in the form
\begin{equation}
\psi(t)
=\sum_{n l m k}
c_{nlmk}\,
e^{-i\omega_n t}\,
\psi_{nlmk}(\sigma,\zeta)\,
f_{nk}(\varepsilon)\,.
\label{eq:Q23}
\end{equation}
Here the conditions \eqref{eq:Q6}, \eqref{eq:Q7} impose for the basis functions the constraints
\begin{align}
\label{eq:Q24}
\hat D_Q \psi_{nlmk}(\sigma,\zeta)&=0\,,
\\
\left(\hat L^2+\hat B^2\right)\psi_{nlmk}(\sigma,\zeta)&=
(n^2-1)\psi_{nlmk}(\sigma,\zeta)\,, \nn
\end{align}
and 
\begin{align}
    &\mathcal{H} \,\psi_{nlmk} (\sigma, \zeta) = -k\,\psi_{nlmk} (\sigma, \zeta) \,, \nn \\
    &\varepsilon^{-3} f_{nk} (\varepsilon) = \varepsilon_k^{-3} (n) f_{nk} (\varepsilon)\,,
    \label{eq:Q25}
\end{align}
with $\mathcal{H}$ given by eq.~\eqref{eq:Q8}.
Up to normalisation eq.~\eqref{eq:Q25} implies $f_{nk} (\varepsilon) \sim \delta^{\frac{1}{2}}(\varepsilon - \varepsilon_k (n))$.

\indent These conditions imply indeed
\begin{align}
\hat C_Q \psi
&=
-\frac{2}{(\kappa\varepsilon)^3}\,
\mathcal{H}\left(\hat L^2+\hat B^2+1\right)\psi
=
-\frac{2n^2}{(\kappa\varepsilon)^3} \mathcal{H}\,\psi \nn \\
&=
\frac{2n^2k}{(\kappa\varepsilon)^3}\psi
=
\frac{2n^2k}{\bigl(\kappa\varepsilon_k(n)\bigr)^3}\psi
=
\psi\,.
\label{eq:Q26}
\end{align}
For the angular momentum the basis functions obey
\begin{align}
&\hat L^2 \psi_{nlmk}(\sigma,\zeta)
=
l(l+1)\psi_{nlmk}(\sigma,\zeta)\,,
\nn \\
&\hat L_3 \psi_{nlmk}(\sigma,\zeta)
=
m\,\psi_{nlmk}(\sigma,\zeta)\,.
\label{eq:Q27}
\end{align}
The basis functions are normalized according to
\begin{align}
&\int_{\sigma,\zeta}
\psi^{*}_{n'l'm'k'}\,
\psi_{nlmk}
=
\delta_{n'n}\delta_{l'l}\delta_{m'm}\delta_{k'k}\,,
\nn \\
&\int_{\varepsilon}
f^{*}_{nk'}\,f_{nk}
=\delta_{k'k}\,.
\label{eq:Q28}
\end{align}
where $\int_{\varepsilon}$ contains the measure factor according to $\int_{zv} = \int_{\sigma, \zeta} \int_{\varepsilon}$.
The normalisation of $\psi(t)$ requires then
\begin{equation}
\sum_{nlmk}|c_{nlmk}|^2=1\,.
\label{eq:Q29}
\end{equation}
The step from the extended hydrogen subsystem to the quantum hydrogen subsystem proceeds by a subtrace of the density matrix which sums over the values of $k$.

\subsection*{Subsystem observables}

\indent Two questions arise at this level. 
The first concerns the classification of the observables compatible with the reduced quantum system. 
The second asks what singles out the constraints \eqref{eq:Q6}, \eqref{eq:Q7}. 
Observables which are compatible with the reduced quantum system are represented by operators $\hat A$ which commute with $\hat D_Q$ and $\hat C_Q$,
\begin{equation}
[\hat A,\hat D_Q]=0\,,
\quad
[\hat A,\hat C_Q]=0\,.
\label{eq:Q18}
\end{equation}
These include all hermitian operators which are constructed of $\hat B_k$, $\hat L_k$ and $H$. 
All those operators commute with $H$, such that their expectation values do not change with time. 

\indent Let us consider general observables which only involve the scale invariant variables $\sigma$, $\zeta$ and derivatives with respect to those variables. 
The dimensionful observables obtain by rescaling with appropriate powers of $\kappa$. 
An observable of this type which is compatible with the reduced subsystem acts linearly on the basis functions $\psi_{nlmk} f_{nk}$,
\begin{equation}
\hat A\,
\psi_{nlmk} f_{nk}
=
\sum_{n'l'm'k'}
\psi_{n'l'm'k'} f_{n'k'}\,
\tilde A_{n'l'm'k',\,nlmk}\,.
\label{eq:Q30}
\end{equation}
Its expectation value is given by
\begin{align}
\langle A\rangle
&=\int_{z,v}
\psi^{*}(t)\hat A\,\psi(t)
=
\operatorname{tr}\bigl(\tilde A\,\tilde\rho(t)\bigr) \nn \\
&=\sum_{n'l'm'k'}\sum_{nlmk} \tilde A_{n'l'm'k',\,nlmk} \,\tilde \rho_{nlmk,\,n'l'm'k'}(t)\,,
\label{eq:Q31}
\end{align}
with density matrix elements
\begin{equation}
\tilde\rho_{nlmk,\,n'l'm'k'}(t)
=
c_{nlmk}\,c^{*}_{n'l'm'k'}\,
\exp\bigl(i(\omega_{n'}-\omega_n)t\bigr)\,.
\label{eq:Q32}
\end{equation}

\indent We are particularly interested in observables which are insensitive to the "winding number" $k$,
\begin{equation}
\tilde A_{nlmk,\,n'l'm'k'}
=
A_{nlm,\,n'l'm'}\,\delta_{kk'}\,.
\label{eq:Q33}
\end{equation}
Their expectation values can be computed from the reduced density matrix $\rho$,
\begin{align}
&\langle A\rangle = \operatorname{tr}(A\rho)\,,
\nn \\
&\rho_{nlm,\,n'l'm'}
=
\sum_k \tilde\rho_{nlmk,\,n'l'm'k}\,.
\label{eq:Q34}
\end{align}
This is the density matrix for the quantum hydrogen subsystem.
The wave functions for pure states obtain in the usual way if $\rho$ obeys the ``pure state condition'' $\rho^2 = \rho$.

\indent For the standard quantum hydrogen atom (without electron spin) every observable is specified uniquely by its representation $A_{nlm,\,n'l'm'}$ in the basis of quantum numbers $(nlm)$.
As a consequence, we can find for each such quantum observable an associated observable for the probabilistic classical particle, defined by eq.~\eqref{eq:Q33}. 
For all these quantum observables the time evolution of expectation values (including correlations) is the same for the quantum electron and the classical electron. 
This follows since the density matrix has the same time evolution for the quantum electron and the classical electron
\begin{equation}
\rho (t) = \exp \bigl(i(\omega_{n'}-\omega_n)t\bigr)\,\rho(0)\,.
\label{eq:Q35}
\end{equation}
The subsystem based on the observables \eqref{eq:Q30}, \eqref{eq:Q33} for the probabilistic classical particle is precisely the same as for the spinless quantum electron.
In particular, the operators for the quantum position and momentum are defined by their matrix values in the $(nlm)$ basis.
They differ from the ones used for free particles and the harmonic oscillator.

\indent The explicit construction of the operator $\hat A$ for a given quantum
observable $A_{nlm,\,n'l'm'}$ needs an explicit form of the basis functions $\psi_{nlmk}(\sigma,\zeta)$. 
This may be rather involved in practice and we have not addressed this issue in the present note. 
It follows from the completeness of the basis $\psi_{nlmk}$ that for every
observable obeying eq.~\eqref{eq:Q33} one operator exists which involves only $\sigma$ and $\zeta$ and derivatives with respect to these variables. 
This shows, in principle, that the map between the quantum electron and the subsystem of the classical probabilistic particle is complete. 
Both systems are equivalent.

\subsection*{Quantum in the sky}

\indent We may apply the quantum constraints \eqref{eq:Q6}, \eqref{eq:Q7} to dust particles in the gravitational field of a star. 
With $m$ the mass of the particle eq.~\eqref{eq:C1A} yields for the physical frequency $\omega_{\text{ph}}=1/T$,
\begin{equation}
\omega_{ph}
=
m\omega
=
\frac{m c^2}{128\pi^2 n^2}
\left(\frac{M_s m}{M^2}\right)^2\,.
\label{eq:Q36}
\end{equation}
For small $n$ this frequency is tiny. 
For ordinary dust on astrophysical scales huge values of $n^2$ are needed. 
This corresponds to the classical limit for the quantum system, for which the discreteness of $n$ can be neglected. 

\indent We can define different quantum subsystems by employing a different eigenvalue of $\hat C_Q$ in eq.~\eqref{eq:Q7},
\begin{equation}
\hat C_Q \psi = \gamma \psi\,.
\label{eq:Q37}
\end{equation}
This still induces a subsystem with the same symmetry structure as the hydrogen atom, but with a different scale for the discrete frequencies
\begin{equation}
\omega = -\gamma\,\frac{\kappa^2}{2n^2}\,.
\label{eq:Q38}
\end{equation}
For example, we may impose a scale invariant constraint
\begin{equation}
\gamma = \frac{2\eta}{(\kappa\epsilon)^3}\,,
\quad
\omega = -\frac{\eta}{\kappa n^2} \varepsilon^{-3} = \frac{\eta}{n^2}\,\omega^{(\text{cl})}\,,
\label{eq:Q39}
\end{equation}
with $\omega^{(\text{cl})}$ the classical frequency \eqref{eq:C2D}.
Choosing $\eta=n_{\max}^2$ this subsystem contains a finite number of discrete energy levels $n=1,\dots,n_{\max}$. 
(For $n>n_{\max}$ the frequency is smaller than $\omega^{(\text{cl})}$, such that no periodic wave function exists.)

\indent More generally, a restriction to eigenstates of suitable statistical observables defines a subsystem for which the frequencies can be discrete.
The restriction to a fixed classical energy $E^{(\text{cl})}=\bar E$ allows for frequencies $k\,\omega^{(\text{cl})}$ with integer $k$. 
We have seen that different subsystems and different spectra of the reduced Hamiltonian are possible. 
For example, we may define a subsystem for a rotating disk of dust. 
The constraints,
\begin{equation}
\hat L^{(\text{cl})}_1\psi=\hat L^{(\text{cl})}_2\psi=0\,,
\label{eq:Q40}
\end{equation}
fix the direction of the classical angular momentum to the direction $3$, without assuming a fixed value for $L^{(\text{cl})}_3$ or $L^{(\text{cl})2}$. 
Rotation invariance around the $3$-axis can be implemented by
\begin{equation}
\hat L_3\psi=0\,,
\label{eq:Q41}
\end{equation}
and we may require parity symmetry
\begin{equation}
P\,\psi=\psi\,.
\label{eq:Q42}
\end{equation}
With these conditions we may investigate given $SO(4)$-representations
\begin{equation}
\hat J_+^2\psi=\hat J_-^2\psi=\frac{1}{4}(n^2-1)\psi\,.
\label{eq:Q43}
\end{equation}
These conditions remain preserved during the time evolution. 
The quantum formulation is a powerful tool for the investigation of the dynamics of such a system.

\indent It is an interesting question if the discreteness of quantum mechanics could be observed in some macroscopic
observables, for example in the probability distribution of the distance of planets to the star.
This could indicate particular stability properties of the quantum subsystem with respect to perturbations.

\section{Periodic time evolution}
\label{sec:XI}

\indent A characteristic feature of quantum mechanics is the periodic time evolution. 
This occurs for generic systems for the eigenstates of the quantum Hamiltonian $H$. 
For the probability distribution of a classical particle a periodic time evolution is expected for the harmonic potential. 
Point particles oscillate with the same frequency independently of their initial position and momentum. 
One is therefore not surprised to find probability distributions which oscillate with this frequency.
For an anharmonic potential a periodic evolution of the probability distribution occurs for particular wave functions. 
Point particles with different initial positions and momenta oscillate with different periods.
General smooth wave functions are not periodic due to this ``dephasing''. 
Periodicity occurs if one restricts the phase space to a suitable compact subspace.
For the example of a one-dimensional particle in an anharmonic potential all trajectories with a fixed classical energy $\bar E_1$ have the same period. 
Wave functions $q = a(x)\,\delta^{1/2}(E_{cl}-\bar E_1)$ oscillate with the frequency $\omega(\bar E_1)$ associated to $\bar E_1$, independently of the initial probability distribution for positions. 
There are also wave functions oscillating with $2\omega(\bar E_1)$.
Furthermore,  a classical energy $\bar E_2$ with $\omega(\bar E_2)=2\omega(\bar E_1)$ may exist. 
The wave functions which are periodic with period $2\omega(\bar E_1)$ can then be superpositions of components with classical energies $\bar E_1$ and $\bar E_2$.
For a particle in three dimensions this issue can become rather complex. 
Stable quantum subsystems with periodic wave functions can be found if there exist compact subspaces of phase space which are closed under the evolution, and if the Hamiltonian restricted to this subspace is bounded.

\subsection*{Eigenstates of $H^2$}

\indent The periodic evolution is characteristic for eigenfunctions of the Hamiltonian. 
We have seen that the only eigenfunction of $H$ which is
compatible with the selection rule \eqref{eq:5} has zero energy and is static.
Periodic wave functions can be obtained, nevertheless, by superpositions of eigenfunctions with positive and negative energy.
Consider the real Hamiltonian \eqref{eq:6} and the pair of eigenfunctions
$\varphi_n(z,s)$ and $\varphi_n(z,-s)$,
\begin{align}
&H\varphi_n(z,s)=\omega_n\varphi_n(z,s)\,,
\nn \\
&H\varphi_n(z,-s)=-\omega_n\varphi_n(z,-s)\,.
\label{eq:P1}
\end{align}
For simplicity we first take $\varphi_n(z,s)$ real. 
A solution of the Schr\"odinger equation \eqref{eq:6} which is compatible with the selection rule reads
\begin{align}
\tilde\varphi_n(t;z,s)
&=
\frac{1}{\sqrt{2}}
\left\{
\varphi_n(z,s)e^{-i\omega_n t}
+
\varphi_n(z,-s)e^{i\omega_n t}
\right\} \nn \\
&=
\tilde\varphi_n^{*}(t;z,-s)\,.
\label{eq:P2}
\end{align}
The wave function \eqref{eq:P2} is an eigenfunction of the operator $H^2$ with eigenvalue $\omega_n^2$. 
We can  generalize the concept of eigenfunctions of $H$ to eigenfunctions of $H^2$. 
They will induce probability distributions which oscillate with period $\pi/\omega_n$. 
Also the superposition of eigenstates with different eigenvalues $\omega_1^2$ and $\omega_2^2$ will induce a periodic time evolution.
The real wave function $q_n(t;z,v)$ corresponding to eq.~\eqref{eq:P2} is given by the inverse of the Fourier transform \eqref{eq:4} of $\tilde\varphi_n$,
\begin{align}
q_n(t;z,v)
&=
\sqrt{2}\,
\Re\left[
\int_s
\exp{\{-i(sv+\omega_n t)\}}\,
\varphi_n(z,s)
\right] \nn \\
&=
\sqrt{2}\int_s
\cos\bigl(sv+\omega_n t\bigr)\,
\varphi_n(z,s)\,.
\label{eq:P3}
\end{align}

\indent More generally, in terms of the complex Fourier transform of $\varphi_n$,
\begin{align}
\chi_n(z,v)
&=
\int_s e^{-isv}\,\varphi_n(z,s)\,ds \nn \\
&=
A(z,v)\exp\bigl(i\alpha(z,v)\bigr)\,,
\label{eq:P4}
\end{align}
with real $A$ and $\alpha$, one has
\begin{align}
q_n(t;z,v)
&=
\frac{1}{\sqrt{2}}
\left\{
e^{-i\omega_n t}\chi_n(z,v)
+
e^{i\omega_n t}\chi_n^{*}(z,v)
\right\} \nn \\
&=
\sqrt{2}\,A(z,v)\cos\bigl(\omega_n t-\alpha(z,v)\bigr)\,.
\label{eq:P5}
\end{align}
The corresponding probability distribution oscillates with period $T=\frac{\pi}{\omega_n}$,
\begin{equation}
w_n
=
q_n^2
=
2\,|\chi_n(z,v)|^2\,
\cos^2\!\bigl(\omega_n t-\alpha(z,v)\bigr)\,.
\label{eq:P6}
\end{equation}
Here the eigenfunctions are normalized according to
\begin{equation}
\int_{z,s} |\varphi_n(z,s)|^2 = 1\,,
\quad
\int_{z,v} |\chi_n(z,v)|^2 = 1\,.
\label{eq:P7}
\end{equation}
For a harmonic potential the lowest positive non-zero eigenvalue of $H$ is $\omega/2$. 
The probability distribution is periodic with frequency $\omega$, as expected.

\subsection*{Periodic probability distributions for harmonic oscillator}

\indent For a harmonic potential the eigenfunctions of the Hamiltonian $H^{(p)}$ for the subsystem of the quantum particle yield static classical wave functions, cf. eqs.~\eqref{eq:FQ47}, \eqref{eq:FQ48}. 
Periodic classical wave functions and associated probability distributions can be found for superpositions of eigenstates of $H^{(p)}$. 
As an example we consider a superposition of the eigenfunctions $\psi^{(0)}$ and $\psi^{(1)}$ in eqs.~\eqref{eq:FQ46}, \eqref{eq:FQ47A}.
\begin{equation}
\tilde\psi(t,x)
=
\alpha\,\psi^{(0)}(t,x)
+
\beta\,\psi^{(1)}(t,x)\,,
\quad
|\alpha|^2+|\beta|^2=1\,.
\label{eq:P8}
\end{equation}
For the wave function $\tilde\varphi$ of the coupled system of particle and mirror particle this results in
\begin{align}
&\tilde\varphi
=
|\alpha|^2\varphi^{(0)}
+
|\beta|^2\varphi^{(1)}
+
\tilde\varphi_v\,,
\nn \\
&\tilde\varphi_v
=
\left(\frac{2}{\pi}\right)^{1/2}
\omega
\left(
\alpha^*\beta\,x\,e^{-i\omega t}
+
\alpha\beta^*\,y\,e^{i\omega t}
\right)
\nn \\
&\hspace{2cm}\times\exp\left(
-\frac{\omega}{2}(x^2+y^2)
\right)\,.
\label{eq:P9}
\end{align}

\indent Taking $\alpha=\beta=1/\sqrt{2}$ the variable part $\tilde\varphi_v$ takes the form
\begin{equation}
\tilde\varphi_v
=
\frac{\omega}{\sqrt{2\pi}}
\left(
2z\cos(\omega t)-is\sin(\omega t)
\right)
\exp\left(
-\omega\left(z^2+\frac{s^2}{4}\right)
\right)
\label{eq:P10}
\end{equation}
resulting in
\begin{align}
\label{eq:P11}
\tilde q(t;z,v)
&=
\frac{1}{2}
\left(
q^{(0)}(z,v)+q^{(1)}(z,v)
\right)
+
\tilde q_v\,,
\\
\tilde q_v
&=
\sqrt{\frac{\omega}{2}}
\left(
2z\cos(\omega t)+\sin(\omega t)\partial_v
\right)
\,q^{(0)}(z,v)
\nonumber\\
&=
\sqrt{2\omega}
\left(
z\cos(\omega t)
-
\frac{v}{\omega}\sin(\omega t)
\right)
q^{(0)}(z,v)\,. \nn 
\end{align}
One easily verifies that $\tilde q$ obeys the Liouville equation, with
\begin{align}
\label{eq:P12}
\tilde q(t;z,v)
=
\frac{1}{2}
\Bigg[
&\left(
\cos(\omega t)+\sqrt{2\omega}\,z
\right)^2
\\
&+
\left(
\sin(\omega t)-\sqrt{\frac{2}{\omega}}\,v
\right)^2
-
1
\Bigg]
q^{(0)}(z,v)\,, \nn
\end{align}
the corresponding periodic probability distribution $\tilde w$ is easily found.

\indent Wave functions with periodic evolution obtain generally by superpositions of eigenstates of $H^{(p)}$ with different energies $E_{q,1}$ and $E_{q,2}$, with period given by $T=2\pi / (\omega_1 - \omega_2)$.
For the harmonic oscillator the frequencies are $n\omega$, $n$ integer.
Already for a system of a few particles with a general quadratic potential a rather rich spectrum of energy eigenvalues and corresponding oscillation periods can be realized in this way. 

\indent For the harmonic potential the spectrum of $H$ can be inferred from the spectrum of $H^{(p)}$.
If $H^{(p)}$ has eigenvalues $\omega_n>0$, the spectrum of eigenvalues $\bar{\omega}_m$ of $H^{(m)}$ is the same. 
With $ H=H^{(p)}-H^{(m)} $ and no interaction between the particle and the mirror particle, the eigenvalues of $H$ are given by the differences $\omega_n-\bar \omega_m$.
For the one-dimensional harmonic oscillator these differences take the values $n\omega$ with integer $n$. 
The eigenvalues of $H^2$ are largely degenerate, since many energy differences lead to the same $(\omega_n-\bar \omega_m)^2$.
Our example \eqref{eq:P12} corresponds to a superposition of the $\omega_n = 3\omega/2$, $\bar\omega_m =\omega/2$ and $\omega_n = \omega/2$, $\bar\omega_m =3\omega/2$.

\subsection*{Periodic observables}

\indent The spectrum of $H$, or the equivalent Liouville operator, is well studied \cite{MMS, AME, KPM, GIA}. 
For the simple case of the anharmonic potential in one dimension it is continuous. 
As discussed above, the eigenfunctions have a sharp classical energy, or are superpositions of such functions.
For the anharmonic potential one may ask if there exist probability distributions for which the behavior of a restricted set of observables based on the quantum position and momentum $x$ and $p$ or similar quantities is periodic. 
This requires no longer exact periodicity in the full phase space. 
Simple arguments based on the trajectories of point particles do no longer apply.
Even if exact periodicity in $x$ and $p$ is not realized, one still expects for a small deviation from the harmonic potential $d$ an impact of the neighbouring periodic solutions for the harmonic potential.
Starting with certain states close to the energy eigenstates of the quantum particle in a harmonic potential, one will find an approximate oscillating behavior for a relatively long time.
In this respect the solutions of the quantum particle in a harmonic potential have special properties among the more general periodic solutions \cite{DW, KAN, MEZI, ACKA, CHRU}.
Similarly, the ``hydrogen subsystem'' for the probabilistic classical particle in the Coulomb potential could have particular stability properties for neighboring settings as small deviations from the exact $1/r$-potential, or other perturbations. 
A discussion of these interesting topics is beyond the scope of this note.

\section{Discussion}
\label{sec:XII}

\indent We propose to look at the Liouville equation for the probabilistic description of a classical point particle from a different angle. 
By the use of statistical observables and a classical wave function it becomes conceptually very close to a quantum particle in a potential. 
In contrast to the standard quantum particle the probabilistic classical particle is described by a different Hamiltonian.
The eigenvalues of this Hamiltonian come in pairs with positive and negative sign.
This reflects the possible description in terms of a real classical wave function which is the root of the probability distribution.
By a Fourier transform we obtain a complex Schr\"odinger equation.
This can be interpreted as describing the quantum system of a particle and its mirror particle. 
In important particular cases, as the particle in a harmonic potential or a free particle, the mirror particle decouples.
The Hamiltonian for the particle takes then precisely the form of the Hamiltonian in the Schr\"odinger equation for a quantum particle.
We can define a subsystem for which the particle is identified with the quantum particle.
Quantum observables are the ones compatible with this subsystem.
All expectation values of the quantum observables for the classical particle take precisely the same values as for a standard quantum particle.

\indent The mirror particle is related to the quantum particle by time reversal.
For the harmonic potential there is no interaction between the mirror particle and the quantum particle. 
In the more general case the interaction between the mirror particle and the quantum particle modifies the evolution of the quantum density matrix for the classical particle as compared to the one of the standard quantum particle.
It is often believed that only for the harmonic potential an equivalence with a quantum system can be established. This is not the case. 
The coupled system of particle and mirror particle is for an arbitrary potential
a genuine quantum system, with all conceptual
properties as non-commuting operators for suitable observables, and a
Schr\"odinger equation or von Neumann equation with a hermitean Hamiltonian.
All quantum laws apply. 
The particular property of the harmonic potential is the decoupling of the particle and the mirror particle. 
This also holds for the free particle.
If one can impose appropriate boundary conditions which are consistent with a closed evolution of the quantum subsystem, the free probabilistic classical particle can account for the interference in the double-slit experiment.

\indent There are different ways to define statistical observables for position and momentum which correspond to the quantum observables for a suitable subsystem.
For the Coulomb potential we have found a quantum subsystem which accounts for all features of the quantum mechanics for the hydrogen atom with a spinless electron.
The associated quantum observables for position and momentum are constructed formally, but we have not attempted to find their explicit expression in terms of the phase space variables and derivatives with respect to these variables.
All these examples give support to the claim that many quantum systems can emerge as subsystems of classical statistical systems \cite{CWQMCS1}, \cite{CWQMCS2}, \cite{CWQMCS3}.

\indent The quantum particles in nature should be seen as the excitations of a vacuum in quantum field theory. 
As compared to the single classical particle discussed here this is a much more complex system with infinitely many degrees of freedom. 
Our formalism applies to classical field theories as well. 
For classical field equations with two time derivatives the variable $z$ corresponds to $\varphi(x)$, and $v$ to $\partial_t \varphi(x)$. 
For a field theory with local interaction the Liouville equation has the same structure as for the classical particle discussed in the present note. 
A free field theory is the analogue of the harmonic potential.
In this case the force term is given by the mass term and the Laplacian in space. 
Here the "mass term" can be inhomogeneous, which can lead to a space-dependent potential for the particle excitations.
It can be shown that a probabilistic classical field theory for a complex scalar field with space-dependent mass term admits a single-particle subsystem which describes a quantum particle in an arbitrary potential \cite{CWTA}.

\indent The infinitely many degrees of freedom of a classical field theory provide for additional flexibility for the realisation of quantum subsystems for single particles.
They also account naturally for the presence of antiparticles.
It is remarkable that even classical systems with a few degrees of freedom, as a point particle in a harmonic or Coulomb potential, can fully describe a subsystem for a quantum particle.
It is an interesting question if some of the characteristic quantum features can be realized in a macroscopic setting, either by dedicated experiments or by observation of systems realized in nature.

\indent Our approach of a quantum description of classical statistical systems can be extended to many problems.
The probabilistic setting for any deterministic evolution equation admits a quantum description.
This covers, in particular, the wide area of transport equations \cite{CWTE}.
Also quantum fermions can be described by the classical statistics of generalized Ising models \cite{CWF1, CWF2, ELZE, CWF3}.

\indent At the present stage it is no longer a question if quantum systems can emerge from classical probabilistic systems. 
The issue is rather if the \textit{observed} quantum systems can be obtained from a classical probabilistic setting.

\nocite{*}
\bibliography{refs}

\begin{thebibliography}{50}%
\makeatletter
\providecommand \@ifxundefined [1]{%
 \@ifx{#1\undefined}
}%
\providecommand \@ifnum [1]{%
 \ifnum #1\expandafter \@firstoftwo
 \else \expandafter \@secondoftwo
 \fi
}%
\providecommand \@ifx [1]{%
 \ifx #1\expandafter \@firstoftwo
 \else \expandafter \@secondoftwo
 \fi
}%
\providecommand \natexlab [1]{#1}%
\providecommand \enquote  [1]{``#1''}%
\providecommand \bibnamefont  [1]{#1}%
\providecommand \bibfnamefont [1]{#1}%
\providecommand \citenamefont [1]{#1}%
\providecommand \href@noop [0]{\@secondoftwo}%
\providecommand \href [0]{\begingroup \@sanitize@url \@href}%
\providecommand \@href[1]{\@@startlink{#1}\@@href}%
\providecommand \@@href[1]{\endgroup#1\@@endlink}%
\providecommand \@sanitize@url [0]{\catcode `\\12\catcode `\$12\catcode
  `\&12\catcode `\#12\catcode `\^12\catcode `\_12\catcode `\%12\relax}%
\providecommand \@@startlink[1]{}%
\providecommand \@@endlink[0]{}%
\providecommand \url  [0]{\begingroup\@sanitize@url \@url }%
\providecommand \@url [1]{\endgroup\@href {#1}{\urlprefix }}%
\providecommand \urlprefix  [0]{URL }%
\providecommand \Eprint [0]{\href }%
\providecommand \doibase [0]{https://doi.org/}%
\providecommand \selectlanguage [0]{\@gobble}%
\providecommand \bibinfo  [0]{\@secondoftwo}%
\providecommand \bibfield  [0]{\@secondoftwo}%
\providecommand \translation [1]{[#1]}%
\providecommand \BibitemOpen [0]{}%
\providecommand \bibitemStop [0]{}%
\providecommand \bibitemNoStop [0]{.\EOS\space}%
\providecommand \EOS [0]{\spacefactor3000\relax}%
\providecommand \BibitemShut  [1]{\csname bibitem#1\endcsname}%
\let\auto@bib@innerbib\@empty
\bibitem [{\citenamefont {Wetterich}(2010{\natexlab{a}})}]{CWQP1}%
  \BibitemOpen
  \bibfield  {author} {\bibinfo {author} {\bibfnamefont {C.}~\bibnamefont
  {Wetterich}},\ }\bibfield  {title} {\bibinfo {title} {Quantum particles from
  classical probabilities in phase space},\ }\href@noop {} {\bibfield
  {journal} {\bibinfo  {journal} {arXiv:1003.0772 [quant-ph]}\ } (\bibinfo
  {year} {2010}{\natexlab{a}})}\BibitemShut {NoStop}%
\bibitem [{\citenamefont {Wetterich}(2010{\natexlab{b}})}]{CWQP2}%
  \BibitemOpen
  \bibfield  {author} {\bibinfo {author} {\bibfnamefont {C.}~\bibnamefont
  {Wetterich}},\ }\bibfield  {title} {\bibinfo {title} {Quantum particles from
  coarse grained classical probabilities in phase space},\ }\href@noop {}
  {\bibfield  {journal} {\bibinfo  {journal} {arXiv:1003.3351 [quant-ph]}\ }
  (\bibinfo {year} {2010}{\natexlab{b}})}\BibitemShut {NoStop}%
\bibitem [{\citenamefont {{Wetterich}}(2018)}]{CWQFC}%
  \BibitemOpen
  \bibfield  {author} {\bibinfo {author} {\bibfnamefont {C.}~\bibnamefont
  {{Wetterich}}},\ }\bibfield  {title} {\bibinfo {title} {{Quantum formalism
  for classical statistics}},\ }\href
  {https://doi.org/10.1016/j.aop.2018.03.022} {\bibfield  {journal} {\bibinfo
  {journal} {Annals of Physics}\ }\textbf {\bibinfo {volume} {393}},\ \bibinfo
  {pages} {1} (\bibinfo {year} {2018})},\ \Eprint
  {https://arxiv.org/abs/1706.01772} {arXiv:1706.01772 [quant-ph]} \BibitemShut
  {NoStop}%
\bibitem [{\citenamefont {Koopman}(1931)}]{KOOP}%
  \BibitemOpen
  \bibfield  {author} {\bibinfo {author} {\bibfnamefont {B.~O.}\ \bibnamefont
  {Koopman}},\ }\bibfield  {title} {\bibinfo {title} {Hamiltonian systems and
  transformation in hilbert space},\ }\href
  {https://doi.org/10.1073/pnas.17.5.315} {\bibfield  {journal} {\bibinfo
  {journal} {Proceedings of the National Academy of Sciences}\ }\textbf
  {\bibinfo {volume} {17}},\ \bibinfo {pages} {315} (\bibinfo {year}
  {1931})}\BibitemShut {NoStop}%
\bibitem [{\citenamefont {v.~Neumann}(1932)}]{VNEU}%
  \BibitemOpen
  \bibfield  {author} {\bibinfo {author} {\bibfnamefont {J.}~\bibnamefont
  {v.~Neumann}},\ }\bibfield  {title} {\bibinfo {title} {Zur operatorenmethode
  in der klassischen mechanik},\ }\href {http://www.jstor.org/stable/1968537}
  {\bibfield  {journal} {\bibinfo  {journal} {Annals of Mathematics}\ }\textbf
  {\bibinfo {volume} {33}},\ \bibinfo {pages} {587} (\bibinfo {year}
  {1932})}\BibitemShut {NoStop}%
\bibitem [{\citenamefont {Mauro}(2002)}]{MAU}%
  \BibitemOpen
  \bibfield  {author} {\bibinfo {author} {\bibfnamefont {D.}~\bibnamefont
  {Mauro}},\ }\bibfield  {title} {\bibinfo {title} {On koopman--von neumann
  waves},\ }\href {https://doi.org/10.1142/S0217751X02009680} {\bibfield
  {journal} {\bibinfo  {journal} {International Journal of Modern Physics A}\
  }\textbf {\bibinfo {volume} {17}},\ \bibinfo {pages} {1301} (\bibinfo {year}
  {2002})},\ \Eprint {https://arxiv.org/abs/quant-ph/0105112}
  {arXiv:quant-ph/0105112 [quant-ph]} \BibitemShut {NoStop}%
\bibitem [{\citenamefont {Gozzi}\ and\ \citenamefont {Mauro}(2004)}]{GOMA}%
  \BibitemOpen
  \bibfield  {author} {\bibinfo {author} {\bibfnamefont {E.}~\bibnamefont
  {Gozzi}}\ and\ \bibinfo {author} {\bibfnamefont {D.}~\bibnamefont {Mauro}},\
  }\bibfield  {title} {\bibinfo {title} {On koopman--von neumann waves ii},\
  }\href {https://doi.org/10.1142/S0217751X04017872} {\bibfield  {journal}
  {\bibinfo  {journal} {International Journal of Modern Physics A}\ }\textbf
  {\bibinfo {volume} {19}},\ \bibinfo {pages} {1475} (\bibinfo {year}
  {2004})},\ \Eprint {https://arxiv.org/abs/quant-ph/0306029}
  {arXiv:quant-ph/0306029 [quant-ph]} \BibitemShut {NoStop}%
\bibitem [{\citenamefont {Bondar}\ \emph {et~al.}(2012)\citenamefont {Bondar},
  \citenamefont {Cabrera}, \citenamefont {Lompay}, \citenamefont {Ivanov},\
  and\ \citenamefont {Rabitz}}]{BON}%
  \BibitemOpen
  \bibfield  {author} {\bibinfo {author} {\bibfnamefont {D.~I.}\ \bibnamefont
  {Bondar}}, \bibinfo {author} {\bibfnamefont {R.}~\bibnamefont {Cabrera}},
  \bibinfo {author} {\bibfnamefont {R.~R.}\ \bibnamefont {Lompay}}, \bibinfo
  {author} {\bibfnamefont {M.~Y.}\ \bibnamefont {Ivanov}},\ and\ \bibinfo
  {author} {\bibfnamefont {H.~A.}\ \bibnamefont {Rabitz}},\ }\bibfield  {title}
  {\bibinfo {title} {Operational dynamic modeling transcending quantum and
  classical mechanics},\ }\href
  {https://doi.org/10.1103/PhysRevLett.109.190403} {\bibfield  {journal}
  {\bibinfo  {journal} {Physical Review Letters}\ }\textbf {\bibinfo {volume}
  {109}},\ \bibinfo {pages} {190403} (\bibinfo {year} {2012})},\ \Eprint
  {https://arxiv.org/abs/1107.5139} {arXiv:1107.5139 [quant-ph]} \BibitemShut
  {NoStop}%
\bibitem [{\citenamefont {{Piasecki}}(2021)}]{PIAS}%
  \BibitemOpen
  \bibfield  {author} {\bibinfo {author} {\bibfnamefont {D.}~\bibnamefont
  {{Piasecki}}},\ }\bibfield  {title} {\bibinfo {title} {{Introduction to
  Koopman-von Neumann Mechanics}},\ }\href@noop {} {\bibfield  {journal}
  {\bibinfo  {journal} {arXiv e-prints}\ ,\ \bibinfo {eid} {arXiv:2112.05619}}
  (\bibinfo {year} {2021})}\BibitemShut {NoStop}%
\bibitem [{\citenamefont {Gozzi}\ and\ \citenamefont {Mauro}(2002)}]{GOMA2}%
  \BibitemOpen
  \bibfield  {author} {\bibinfo {author} {\bibfnamefont {E.}~\bibnamefont
  {Gozzi}}\ and\ \bibinfo {author} {\bibfnamefont {D.}~\bibnamefont {Mauro}},\
  }\bibfield  {title} {\bibinfo {title} {Minimal coupling in koopman–von
  neumann theory},\ }\href
  {https://doi.org/https://doi.org/10.1006/aphy.2001.6206} {\bibfield
  {journal} {\bibinfo  {journal} {Annals of Physics}\ }\textbf {\bibinfo
  {volume} {296}},\ \bibinfo {pages} {152} (\bibinfo {year}
  {2002})}\BibitemShut {NoStop}%
\bibitem [{\citenamefont {Klein}(2018)}]{KLE}%
  \BibitemOpen
  \bibfield  {author} {\bibinfo {author} {\bibfnamefont {U.}~\bibnamefont
  {Klein}},\ }\bibfield  {title} {\bibinfo {title} {From koopman–von neumann
  theory to quantum theory},\ }\href
  {https://doi.org/10.1007/s40509-017-0113-2} {\bibfield  {journal} {\bibinfo
  {journal} {Quantum Studies: Mathematics and Foundations}\ }\textbf {\bibinfo
  {volume} {5}},\ \bibinfo {pages} {219} (\bibinfo {year} {2018})},\ \Eprint
  {https://arxiv.org/abs/1705.07427} {arXiv:1705.07427 [math.DS]} \BibitemShut
  {NoStop}%
\bibitem [{\citenamefont {Mauroy}\ \emph {et~al.}(2020)\citenamefont {Mauroy},
  \citenamefont {Mezi{\'c}},\ and\ \citenamefont {Susuki}}]{MMS}%
  \BibitemOpen
  \bibinfo {editor} {\bibfnamefont {A.}~\bibnamefont {Mauroy}}, \bibinfo
  {editor} {\bibfnamefont {I.}~\bibnamefont {Mezi{\'c}}},\ and\ \bibinfo
  {editor} {\bibfnamefont {Y.}~\bibnamefont {Susuki}},\ eds.,\ \href
  {https://doi.org/10.1007/978-3-030-35713-9} {\emph {\bibinfo {title} {The
  Koopman Operator in Systems and Control: Concepts, Methodologies, and
  Applications}}},\ \bibinfo {series} {Lecture Notes in Control and Information
  Sciences}, Vol.\ \bibinfo {volume} {484}\ (\bibinfo  {publisher} {Springer},\
  \bibinfo {address} {Cham},\ \bibinfo {year} {2020})\BibitemShut {NoStop}%
\bibitem [{\citenamefont {Arbabi}\ and\ \citenamefont {Mezi\'c}(2017)}]{AME}%
  \BibitemOpen
  \bibfield  {author} {\bibinfo {author} {\bibfnamefont {H.}~\bibnamefont
  {Arbabi}}\ and\ \bibinfo {author} {\bibfnamefont {I.}~\bibnamefont
  {Mezi\'c}},\ }\bibfield  {title} {\bibinfo {title} {Ergodic theory, dynamic
  mode decomposition and computation of spectral properties of the koopman
  operator},\ }\href@noop {} {\bibfield  {journal} {\bibinfo  {journal} {SIAM
  Journal on Applied Dynamical Systems}\ }\textbf {\bibinfo {volume} {16}},\
  \bibinfo {pages} {2096} (\bibinfo {year} {2017})},\ \Eprint
  {https://arxiv.org/abs/1611.06664} {arXiv:1611.06664 [math.DS]} \BibitemShut
  {NoStop}%
\bibitem [{\citenamefont {Korda}\ \emph {et~al.}(2017)\citenamefont {Korda},
  \citenamefont {Putinar},\ and\ \citenamefont {Mezi{\'c}}}]{KPM}%
  \BibitemOpen
  \bibfield  {author} {\bibinfo {author} {\bibfnamefont {M.}~\bibnamefont
  {Korda}}, \bibinfo {author} {\bibfnamefont {M.}~\bibnamefont {Putinar}},\
  and\ \bibinfo {author} {\bibfnamefont {I.}~\bibnamefont {Mezi{\'c}}},\
  }\bibfield  {title} {\bibinfo {title} {Data‐driven spectral analysis of the
  koopman operator},\ }\href@noop {} {\bibfield  {journal} {\bibinfo  {journal}
  {arXiv:1710.06532}\ } (\bibinfo {year} {2017})}\BibitemShut {NoStop}%
\bibitem [{\citenamefont {{Giannakis}}(2020)}]{GIA}%
  \BibitemOpen
  \bibfield  {author} {\bibinfo {author} {\bibfnamefont {D.}~\bibnamefont
  {{Giannakis}}},\ }\bibfield  {title} {\bibinfo {title} {{Delay-coordinate
  maps, coherence, and approximate spectra of evolution operators}},\
  }\href@noop {} {\bibfield  {journal} {\bibinfo  {journal} {arXiv e-prints}\
  ,\ \bibinfo {eid} {arXiv:2007.02195}} (\bibinfo {year} {2020})}\BibitemShut
  {NoStop}%
\bibitem [{\citenamefont {Chru{\'s}ci{\'n}ski}(2006)}]{CHRU}%
  \BibitemOpen
  \bibfield  {author} {\bibinfo {author} {\bibfnamefont {D.}~\bibnamefont
  {Chru{\'s}ci{\'n}ski}},\ }\bibfield  {title} {\bibinfo {title} {Koopman’s
  approach to dissipation},\ }\href@noop {} {\bibfield  {journal} {\bibinfo
  {journal} {Reports on Mathematical Physics}\ }\textbf {\bibinfo {volume}
  {57}},\ \bibinfo {pages} {319} (\bibinfo {year} {2006})}\BibitemShut
  {NoStop}%
\bibitem [{\citenamefont {Mezi{\'c}}(2022)}]{MEZI}%
  \BibitemOpen
  \bibfield  {author} {\bibinfo {author} {\bibfnamefont {I.}~\bibnamefont
  {Mezi{\'c}}},\ }\bibfield  {title} {\bibinfo {title} {On numerical
  approximations of the koopman operator},\ }\href
  {https://doi.org/10.3390/math10071180} {\bibfield  {journal} {\bibinfo
  {journal} {Mathematics}\ }\textbf {\bibinfo {volume} {10}},\ \bibinfo {pages}
  {1180} (\bibinfo {year} {2022})},\ \Eprint {https://arxiv.org/abs/2009.05883}
  {arXiv:2009.05883 [math.DS]} \BibitemShut {NoStop}%
\bibitem [{\citenamefont {Darling}\ and\ \citenamefont {Widrow}(2024)}]{DW}%
  \BibitemOpen
  \bibfield  {author} {\bibinfo {author} {\bibfnamefont {K.}~\bibnamefont
  {Darling}}\ and\ \bibinfo {author} {\bibfnamefont {L.~M.}\ \bibnamefont
  {Widrow}},\ }\bibfield  {title} {\bibinfo {title} {Linear operator theory of
  phase mixing},\ }\href {https://doi.org/10.1093/mnras/stae1775} {\bibfield
  {journal} {\bibinfo  {journal} {Monthly Notices of the Royal Astronomical
  Society}\ }\textbf {\bibinfo {volume} {533}},\ \bibinfo {pages} {79}
  (\bibinfo {year} {2024})}\BibitemShut {NoStop}%
\bibitem [{\citenamefont {Gay-Balmaz}\ and\ \citenamefont
  {Tronci}(2021)}]{GBT}%
  \BibitemOpen
  \bibfield  {author} {\bibinfo {author} {\bibfnamefont {F.}~\bibnamefont
  {Gay-Balmaz}}\ and\ \bibinfo {author} {\bibfnamefont {C.}~\bibnamefont
  {Tronci}},\ }\bibfield  {title} {\bibinfo {title} {From quantum hydrodynamics
  to koopman wavefunctions i},\ }in\ \href@noop {} {\emph {\bibinfo {booktitle}
  {Geometric Science of Information}}},\ \bibinfo {editor} {edited by\ \bibinfo
  {editor} {\bibfnamefont {F.}~\bibnamefont {Nielsen}}\ and\ \bibinfo {editor}
  {\bibfnamefont {F.}~\bibnamefont {Barbaresco}}}\ (\bibinfo  {publisher}
  {Springer International Publishing},\ \bibinfo {address} {Cham},\ \bibinfo
  {year} {2021})\ pp.\ \bibinfo {pages} {302--310},\ \Eprint
  {https://arxiv.org/abs/2104.13185} {arXiv:2104.13185} \BibitemShut {NoStop}%
\bibitem [{\citenamefont {Gozzi}\ \emph {et~al.}(1989)\citenamefont {Gozzi},
  \citenamefont {Reuter},\ and\ \citenamefont {Thacker}}]{GRT}%
  \BibitemOpen
  \bibfield  {author} {\bibinfo {author} {\bibfnamefont {E.}~\bibnamefont
  {Gozzi}}, \bibinfo {author} {\bibfnamefont {M.}~\bibnamefont {Reuter}},\ and\
  \bibinfo {author} {\bibfnamefont {W.~D.}\ \bibnamefont {Thacker}},\
  }\bibfield  {title} {\bibinfo {title} {Hidden brs invariance in classical
  mechanics. ii},\ }\href {https://doi.org/10.1103/PhysRevD.40.3363} {\bibfield
   {journal} {\bibinfo  {journal} {Phys. Rev. D}\ }\textbf {\bibinfo {volume}
  {40}},\ \bibinfo {pages} {3363} (\bibinfo {year} {1989})}\BibitemShut
  {NoStop}%
\bibitem [{\citenamefont {Kofler}\ and\ \citenamefont {Brukner}(2007)}]{KOBR}%
  \BibitemOpen
  \bibfield  {author} {\bibinfo {author} {\bibfnamefont {J.}~\bibnamefont
  {Kofler}}\ and\ \bibinfo {author} {\bibfnamefont {i.~c.~v.}\ \bibnamefont
  {Brukner}},\ }\bibfield  {title} {\bibinfo {title} {Classical world arising
  out of quantum physics under the restriction of coarse-grained
  measurements},\ }\href {https://doi.org/10.1103/PhysRevLett.99.180403}
  {\bibfield  {journal} {\bibinfo  {journal} {Phys. Rev. Lett.}\ }\textbf
  {\bibinfo {volume} {99}},\ \bibinfo {pages} {180403} (\bibinfo {year}
  {2007})}\BibitemShut {NoStop}%
\bibitem [{\citenamefont {{Krishnan V}}\ \emph {et~al.}(2017)\citenamefont
  {{Krishnan V}}, \citenamefont {{Biswas}},\ and\ \citenamefont
  {{Ghosh}}}]{KRBG}%
  \BibitemOpen
  \bibfield  {author} {\bibinfo {author} {\bibfnamefont {M.}~\bibnamefont
  {{Krishnan V}}}, \bibinfo {author} {\bibfnamefont {T.}~\bibnamefont
  {{Biswas}}},\ and\ \bibinfo {author} {\bibfnamefont {S.}~\bibnamefont
  {{Ghosh}}},\ }\bibfield  {title} {\bibinfo {title} {{Coarse-graining of
  measurement and quantum-to-classical transition in the bipartite scenario}},\
  }\href@noop {} {\bibfield  {journal} {\bibinfo  {journal} {arXiv e-prints}\
  ,\ \bibinfo {eid} {arXiv:1703.00502}} (\bibinfo {year} {2017})}\BibitemShut
  {NoStop}%
\bibitem [{\citenamefont {Gallego}(2022)}]{GAL}%
  \BibitemOpen
  \bibfield  {author} {\bibinfo {author} {\bibfnamefont {M.}~\bibnamefont
  {Gallego}},\ }\bibfield  {title} {\bibinfo {title} {Coarse-graining and the
  quantum-to-classical transition},\ }\href
  {https://doi.org/10.1142/S0217979222300018} {\bibfield  {journal} {\bibinfo
  {journal} {International Journal of Modern Physics B}\ }\textbf {\bibinfo
  {volume} {36}},\ \bibinfo {pages} {2230001} (\bibinfo {year}
  {2022})}\BibitemShut {NoStop}%
\bibitem [{\citenamefont {{Bibak}}\ \emph {et~al.}(2025)\citenamefont
  {{Bibak}}, \citenamefont {{Cepollaro}}, \citenamefont {{Medina S{\'a}nchez}},
  \citenamefont {{Daki{\'c}}},\ and\ \citenamefont {{Brukner}}}]{BICE}%
  \BibitemOpen
  \bibfield  {author} {\bibinfo {author} {\bibfnamefont {F.}~\bibnamefont
  {{Bibak}}}, \bibinfo {author} {\bibfnamefont {C.}~\bibnamefont
  {{Cepollaro}}}, \bibinfo {author} {\bibfnamefont {N.}~\bibnamefont {{Medina
  S{\'a}nchez}}}, \bibinfo {author} {\bibfnamefont {B.}~\bibnamefont
  {{Daki{\'c}}}},\ and\ \bibinfo {author} {\bibfnamefont
  {{\v{C}}.}~\bibnamefont {{Brukner}}},\ }\bibfield  {title} {\bibinfo {title}
  {{The classical limit of quantum mechanics through coarse-grained
  measurements}},\ }\href@noop {} {\bibfield  {journal} {\bibinfo  {journal}
  {arXiv e-prints}\ } (\bibinfo {year} {2025})},\ \Eprint
  {https://arxiv.org/abs/2503.15642} {arXiv:2503.15642 [quant-ph]} \BibitemShut
  {NoStop}%
\bibitem [{\citenamefont {{Berges}}(2015)}]{BER}%
  \BibitemOpen
  \bibfield  {author} {\bibinfo {author} {\bibfnamefont {J.}~\bibnamefont
  {{Berges}}},\ }\bibfield  {title} {\bibinfo {title} {{Nonequilibrium Quantum
  Fields: From Cold Atoms to Cosmology}},\ }\href
  {https://doi.org/10.48550/arXiv.1503.02907} {\bibfield  {journal} {\bibinfo
  {journal} {arXiv e-prints}\ ,\ \bibinfo {eid} {arXiv:1503.02907 [hep-ph]}}
  (\bibinfo {year} {2015})}\BibitemShut {NoStop}%
\bibitem [{\citenamefont {Kandrup}(1998)}]{KAN}%
  \BibitemOpen
  \bibfield  {author} {\bibinfo {author} {\bibfnamefont {H.~E.}\ \bibnamefont
  {Kandrup}},\ }\bibfield  {title} {\bibinfo {title} {Phase mixing in
  time-independent hamiltonian systems},\ }\href
  {https://doi.org/10.1046/j.1365-8711.1998.02063.x} {\bibfield  {journal}
  {\bibinfo  {journal} {Monthly Notices of the Royal Astronomical Society}\
  }\textbf {\bibinfo {volume} {301}},\ \bibinfo {pages} {960} (\bibinfo {year}
  {1998})}\BibitemShut {NoStop}%
\bibitem [{\citenamefont {Man'ko}\ and\ \citenamefont {Marmo}(1999)}]{MAMA}%
  \BibitemOpen
  \bibfield  {author} {\bibinfo {author} {\bibfnamefont {V.~I.}\ \bibnamefont
  {Man'ko}}\ and\ \bibinfo {author} {\bibfnamefont {G.}~\bibnamefont {Marmo}},\
  }\bibfield  {title} {\bibinfo {title} {Alternative commutation relations,
  star products and tomography},\ }\href
  {https://doi.org/10.1238/physica.scripta.60.111} {\bibfield  {journal}
  {\bibinfo  {journal} {Physica Scripta}\ }\textbf {\bibinfo {volume} {60}},\
  \bibinfo {pages} {111} (\bibinfo {year} {1999})},\ \Eprint
  {https://arxiv.org/abs/quant-ph/9903021} {quant-ph/9903021} \BibitemShut
  {NoStop}%
\bibitem [{\citenamefont {Nikoli{\'c}}(2006)}]{NIC}%
  \BibitemOpen
  \bibfield  {author} {\bibinfo {author} {\bibfnamefont {H.}~\bibnamefont
  {Nikoli{\'c}}},\ }\bibfield  {title} {\bibinfo {title} {Classical mechanics
  without determinism},\ }\href@noop {} {\bibfield  {journal} {\bibinfo
  {journal} {Foundations of Physics Letters}\ }\textbf {\bibinfo {volume}
  {19}},\ \bibinfo {pages} {553} (\bibinfo {year} {2006})},\ \Eprint
  {https://arxiv.org/abs/quant-ph/0505143} {arXiv:quant-ph/0505143 [quant-ph]}
  \BibitemShut {NoStop}%
\bibitem [{\citenamefont {Volovich}(2011)}]{VOLO}%
  \BibitemOpen
  \bibfield  {author} {\bibinfo {author} {\bibfnamefont {I.~V.}\ \bibnamefont
  {Volovich}},\ }\bibfield  {title} {\bibinfo {title} {Randomness in classical
  mechanics and quantum mechanics},\ }\href
  {https://doi.org/10.1007/s10701-010-9450-2} {\bibfield  {journal} {\bibinfo
  {journal} {Foundations of Physics}\ }\textbf {\bibinfo {volume} {41}},\
  \bibinfo {pages} {516} (\bibinfo {year} {2011})},\ \Eprint
  {https://arxiv.org/abs/0910.5391} {arXiv:0910.5391 [quant-ph]} \BibitemShut
  {NoStop}%
\bibitem [{\citenamefont {Arsiwalla}\ \emph {et~al.}(2024)\citenamefont
  {Arsiwalla}, \citenamefont {Chester},\ and\ \citenamefont {Kauffman}}]{ACKA}%
  \BibitemOpen
  \bibfield  {author} {\bibinfo {author} {\bibfnamefont {X.~D.}\ \bibnamefont
  {Arsiwalla}}, \bibinfo {author} {\bibfnamefont {D.}~\bibnamefont {Chester}},\
  and\ \bibinfo {author} {\bibfnamefont {L.~H.}\ \bibnamefont {Kauffman}},\
  }\bibfield  {title} {\bibinfo {title} {On the operator origins of classical
  and quantum wave functions},\ }\href
  {https://doi.org/10.1007/s40509-023-00311-6} {\bibfield  {journal} {\bibinfo
  {journal} {Quantum Studies: Mathematics and Foundations}\ }\textbf {\bibinfo
  {volume} {11}},\ \bibinfo {pages} {193} (\bibinfo {year} {2024})},\ \Eprint
  {https://arxiv.org/abs/2211.01838} {arXiv:2211.01838 [math-ph]} \BibitemShut
  {NoStop}%
\bibitem [{\citenamefont {Gozzi}(1988)}]{GOZ1}%
  \BibitemOpen
  \bibfield  {author} {\bibinfo {author} {\bibfnamefont {E.}~\bibnamefont
  {Gozzi}},\ }\bibfield  {title} {\bibinfo {title} {Hidden brs invariance in
  classical mechanics},\ }\href {https://doi.org/10.1016/0370-2693(88)90688-2}
  {\bibfield  {journal} {\bibinfo  {journal} {Phys. Lett. B}\ }\textbf
  {\bibinfo {volume} {201}},\ \bibinfo {pages} {525} (\bibinfo {year}
  {1988})}\BibitemShut {NoStop}%
\bibitem [{\citenamefont {Wetterich}(1997{\natexlab{a}})}]{CWET1}%
  \BibitemOpen
  \bibfield  {author} {\bibinfo {author} {\bibfnamefont {C.}~\bibnamefont
  {Wetterich}},\ }\bibfield  {title} {\bibinfo {title} {Time evolution of
  non-equilibrium effective action},\ }\href
  {https://doi.org/10.1103/PhysRevLett.78.3598} {\bibfield  {journal} {\bibinfo
   {journal} {Physical Review Letters}\ }\textbf {\bibinfo {volume} {78}},\
  \bibinfo {pages} {3598} (\bibinfo {year} {1997}{\natexlab{a}})},\ \Eprint
  {https://arxiv.org/abs/hep-th/9612206} {arXiv:hep-th/9612206 [hep-th]}
  \BibitemShut {NoStop}%
\bibitem [{\citenamefont {Wetterich}(1997{\natexlab{b}})}]{CWET2}%
  \BibitemOpen
  \bibfield  {author} {\bibinfo {author} {\bibfnamefont {C.}~\bibnamefont
  {Wetterich}},\ }\bibfield  {title} {\bibinfo {title} {Non-equilibrium time
  evolution in quantum field theory},\ }\href
  {https://doi.org/10.1103/PhysRevE.56.2687} {\bibfield  {journal} {\bibinfo
  {journal} {Physical Review E}\ }\textbf {\bibinfo {volume} {56}},\ \bibinfo
  {pages} {2687} (\bibinfo {year} {1997}{\natexlab{b}})},\ \Eprint
  {https://arxiv.org/abs/hep-th/9703006} {arXiv:hep-th/9703006 [hep-th]}
  \BibitemShut {NoStop}%
\bibitem [{\citenamefont {Nikoli{\'c}}(2007)}]{NAKL}%
  \BibitemOpen
  \bibfield  {author} {\bibinfo {author} {\bibfnamefont {H.}~\bibnamefont
  {Nikoli{\'c}}},\ }\bibfield  {title} {\bibinfo {title} {Classical mechanics
  as nonlinear quantum mechanics},\ }\href@noop {} {\bibfield  {journal}
  {\bibinfo  {journal} {arXiv:quant-ph/0707.2319}\ } (\bibinfo {year}
  {2007})}\BibitemShut {NoStop}%
\bibitem [{\citenamefont {Wetterich}(2004)}]{CWQMCS1}%
  \BibitemOpen
  \bibfield  {author} {\bibinfo {author} {\bibfnamefont {C.}~\bibnamefont
  {Wetterich}},\ }\bibfield  {title} {\bibinfo {title} {Quantum correlations in
  classical statistics},\ }in\ \href
  {https://doi.org/10.1007/978-3-540-40968-7_14} {\emph {\bibinfo {booktitle}
  {Decoherence and Entropy in Complex Systems}}},\ \bibinfo {series} {Lecture
  Notes in Physics}, Vol.\ \bibinfo {volume} {633},\ \bibinfo {editor} {edited
  by\ \bibinfo {editor} {\bibfnamefont {H.-T.}\ \bibnamefont {Elze}}}\
  (\bibinfo  {publisher} {Springer},\ \bibinfo {address} {Berlin, Heidelberg},\
  \bibinfo {year} {2004})\ pp.\ \bibinfo {pages} {180--195},\ \Eprint
  {https://arxiv.org/abs/quant-ph/0212031} {arXiv:quant-ph/0212031 [quant-ph]}
  \BibitemShut {NoStop}%
\bibitem [{\citenamefont {{Wetterich}}(2016)}]{CWIT}%
  \BibitemOpen
  \bibfield  {author} {\bibinfo {author} {\bibfnamefont {C.}~\bibnamefont
  {{Wetterich}}},\ }\bibfield  {title} {\bibinfo {title} {{Information
  transport in classical statistical systems}},\ }\href
  {https://doi.org/10.48550/arXiv.1611.04820} {\bibfield  {journal} {\bibinfo
  {journal} {arXiv e-prints}\ ,\ \bibinfo {eid} {arXiv:1611.04820
  [cond-mat.stat-mech]}} (\bibinfo {year} {2016})}\BibitemShut {NoStop}%
\bibitem [{\citenamefont {{Wetterich}}(2026)}]{CW26}%
  \BibitemOpen
  \bibfield  {author} {\bibinfo {author} {\bibfnamefont {C.}~\bibnamefont
  {{Wetterich}}},\ }\bibfield  {title} {\bibinfo {title} {{Quantum field theory
  for classical fields}},\ }\href {https://doi.org/10.48550/arXiv.2603.05061}
  {\bibfield  {journal} {\bibinfo  {journal} {arXiv e-prints}\ ,\ \bibinfo
  {eid} {arXiv:2603.05061}} (\bibinfo {year} {2026})}\BibitemShut {NoStop}%
\bibitem [{\citenamefont {Bell}(1964)}]{BEL1}%
  \BibitemOpen
  \bibfield  {author} {\bibinfo {author} {\bibfnamefont {J.~S.}\ \bibnamefont
  {Bell}},\ }\bibfield  {title} {\bibinfo {title} {On the einstein podolsky
  rosen paradox},\ }\href {https://doi.org/10.1103/PhysicsPhysiqueFizika.1.195}
  {\bibfield  {journal} {\bibinfo  {journal} {Physics Physique Fizika}\
  }\textbf {\bibinfo {volume} {1}},\ \bibinfo {pages} {195} (\bibinfo {year}
  {1964})}\BibitemShut {NoStop}%
\bibitem [{\citenamefont {Clauser}\ \emph {et~al.}(1969)\citenamefont
  {Clauser}, \citenamefont {Horne}, \citenamefont {Shimony},\ and\
  \citenamefont {Holt}}]{CHSH}%
  \BibitemOpen
  \bibfield  {author} {\bibinfo {author} {\bibfnamefont {J.~F.}\ \bibnamefont
  {Clauser}}, \bibinfo {author} {\bibfnamefont {M.~A.}\ \bibnamefont {Horne}},
  \bibinfo {author} {\bibfnamefont {A.}~\bibnamefont {Shimony}},\ and\ \bibinfo
  {author} {\bibfnamefont {R.~A.}\ \bibnamefont {Holt}},\ }\bibfield  {title}
  {\bibinfo {title} {Proposed experiment to test local hidden-variable
  theories},\ }\href {https://doi.org/10.1103/PhysRevLett.23.880} {\bibfield
  {journal} {\bibinfo  {journal} {Phys. Rev. Lett.}\ }\textbf {\bibinfo
  {volume} {23}},\ \bibinfo {pages} {880} (\bibinfo {year} {1969})}\BibitemShut
  {NoStop}%
\bibitem [{\citenamefont {Wetterich}(2025)}]{CWPW}%
  \BibitemOpen
  \bibfield  {author} {\bibinfo {author} {\bibfnamefont {C.}~\bibnamefont
  {Wetterich}},\ }\href@noop {} {\emph {\bibinfo {title} {The Probabilistic
  World}}}\ (\bibinfo  {publisher} {Springer Nature},\ \bibinfo {address}
  {Heidelberg},\ \bibinfo {year} {2025})\ \Eprint
  {https://arxiv.org/abs/2011.02867} {arXiv:2011.02867 [quant-ph]} \BibitemShut
  {NoStop}%
\bibitem [{\citenamefont {{Wetterich}}(2024)}]{CWPW2}%
  \BibitemOpen
  \bibfield  {author} {\bibinfo {author} {\bibfnamefont {C.}~\bibnamefont
  {{Wetterich}}},\ }\bibfield  {title} {\bibinfo {title} {{The probabilistic
  world II : Quantum mechanics from classical statistics}},\ }\href@noop {}
  {\bibfield  {journal} {\bibinfo  {journal} {arXiv e-prints}\ ,\ \bibinfo
  {eid} {arXiv:2408.06379}} (\bibinfo {year} {2024})}\BibitemShut {NoStop}%
\bibitem [{\citenamefont {Barnes}(1995)}]{SUBA}%
  \BibitemOpen
  \bibfield  {author} {\bibinfo {author} {\bibfnamefont {I.~S.}\ \bibnamefont
  {Barnes}},\ }\bibfield  {title} {\bibinfo {title} {Semiclassical wavepacket
  propagation in a hydrogen atom},\ }\href
  {https://doi.org/10.1016/0960-0779(95)80061-K} {\bibfield  {journal}
  {\bibinfo  {journal} {Chaos, Solitons \& Fractals}\ }\textbf {\bibinfo
  {volume} {6}},\ \bibinfo {pages} {531} (\bibinfo {year} {1995})},\ \bibinfo
  {note} {complex Systems in Computational Physics}\BibitemShut {NoStop}%
\bibitem [{\citenamefont {Wetterich}(2010{\natexlab{c}})}]{CWQMCS2}%
  \BibitemOpen
  \bibfield  {author} {\bibinfo {author} {\bibfnamefont {C.}~\bibnamefont
  {Wetterich}},\ }\bibfield  {title} {\bibinfo {title} {Probabilistic
  observables, conditional correlations, and quantum physics},\ }\href
  {https://doi.org/10.1002/andp.201010451} {\bibfield  {journal} {\bibinfo
  {journal} {Annalen der Physik}\ }\textbf {\bibinfo {volume} {522}},\ \bibinfo
  {pages} {467} (\bibinfo {year} {2010}{\natexlab{c}})},\ \Eprint
  {https://arxiv.org/abs/0810.0985} {arXiv:0810.0985 [quant-ph]} \BibitemShut
  {NoStop}%
\bibitem [{\citenamefont {Wetterich}(2009)}]{CWQMCS3}%
  \BibitemOpen
  \bibfield  {author} {\bibinfo {author} {\bibfnamefont {C.}~\bibnamefont
  {Wetterich}},\ }\bibfield  {title} {\bibinfo {title} {Emergence of quantum
  mechanics from classical statistics},\ }\href
  {https://doi.org/10.1088/1742-6596/174/1/012008} {\bibfield  {journal}
  {\bibinfo  {journal} {Journal of Physics: Conference Series}\ }\textbf
  {\bibinfo {volume} {174}},\ \bibinfo {pages} {012008} (\bibinfo {year}
  {2009})},\ \Eprint {https://arxiv.org/abs/0811.0927} {arXiv:0811.0927
  [quant-ph]} \BibitemShut {NoStop}%
\bibitem [{\citenamefont {Wetterich}(2026{\natexlab{a}})}]{CWTA}%
  \BibitemOpen
  \bibfield  {author} {\bibinfo {author} {\bibfnamefont {C.}~\bibnamefont
  {Wetterich}},\ }\href {https://doi.org/10.48550/arXiv.2607.11580} {\bibinfo
  {title} {Classical probabilistic realisation of quantum double-slit
  interference}} (\bibinfo {year} {2026}{\natexlab{a}}),\ \bibinfo {note}
  {arXiv preprint},\ \Eprint {https://arxiv.org/abs/2607.11580}
  {arXiv:2607.11580 [quant-ph]} \BibitemShut {NoStop}%
\bibitem [{\citenamefont {Wetterich}(2026{\natexlab{b}})}]{CWTE}%
  \BibitemOpen
  \bibfield  {author} {\bibinfo {author} {\bibfnamefont {C.}~\bibnamefont
  {Wetterich}},\ }\href {https://doi.org/10.48550/arXiv.2605.15969} {\bibinfo
  {title} {Quantum mechanics for classical transport equations}} (\bibinfo
  {year} {2026}{\natexlab{b}}),\ \bibinfo {note} {arXiv preprint},\ \Eprint
  {https://arxiv.org/abs/2605.15969} {arXiv:2605.15969 [quant-ph]} \BibitemShut
  {NoStop}%
\bibitem [{\citenamefont {Wetterich}(2017)}]{CWF1}%
  \BibitemOpen
  \bibfield  {author} {\bibinfo {author} {\bibfnamefont {C.}~\bibnamefont
  {Wetterich}},\ }\bibfield  {title} {\bibinfo {title} {Fermions as generalized
  ising models},\ }\href {https://doi.org/10.1016/j.nuclphysb.2017.02.012}
  {\bibfield  {journal} {\bibinfo  {journal} {Nuclear Physics B}\ }\textbf
  {\bibinfo {volume} {917}},\ \bibinfo {pages} {241} (\bibinfo {year}
  {2017})},\ \Eprint {https://arxiv.org/abs/1612.06695} {arXiv:1612.06695
  [cond-mat.stat-mech]} \BibitemShut {NoStop}%
\bibitem [{\citenamefont {Wetterich}(2022)}]{CWF2}%
  \BibitemOpen
  \bibfield  {author} {\bibinfo {author} {\bibfnamefont {C.}~\bibnamefont
  {Wetterich}},\ }\href {https://doi.org/10.48550/arXiv.2203.14081} {\bibinfo
  {title} {Fermion picture for cellular automata}} (\bibinfo {year} {2022}),\
  \bibinfo {note} {arXiv preprint},\ \Eprint {https://arxiv.org/abs/2203.14081}
  {arXiv:2203.14081 [nlin.CG]} \BibitemShut {NoStop}%
\bibitem [{\citenamefont {Elze}(2025)}]{ELZE}%
  \BibitemOpen
  \bibfield  {author} {\bibinfo {author} {\bibfnamefont {H.-T.}\ \bibnamefont
  {Elze}},\ }\bibfield  {title} {\bibinfo {title} {The dirac equation, mass and
  arithmetic by permutations of automaton states},\ }\href
  {https://doi.org/10.3390/e27040395} {\bibfield  {journal} {\bibinfo
  {journal} {Entropy}\ }\textbf {\bibinfo {volume} {27}},\ \bibinfo {pages}
  {395} (\bibinfo {year} {2025})},\ \Eprint {https://arxiv.org/abs/2504.06883}
  {arXiv:2504.06883 [quant-ph]} \BibitemShut {NoStop}%
\bibitem [{\citenamefont {Wetterich}(2026{\natexlab{c}})}]{CWF3}%
  \BibitemOpen
  \bibfield  {author} {\bibinfo {author} {\bibfnamefont {C.}~\bibnamefont
  {Wetterich}},\ }\bibfield  {title} {\bibinfo {title} {Complex wave functions,
  {CPT} and quantum field theory for classical generalized {Ising} models},\
  }\href {https://doi.org/10.1016/j.nuclphysb.2025.117204} {\bibfield
  {journal} {\bibinfo  {journal} {Nuclear Physics B}\ }\textbf {\bibinfo
  {volume} {1022}},\ \bibinfo {pages} {117204} (\bibinfo {year}
  {2026}{\natexlab{c}})},\ \Eprint {https://arxiv.org/abs/2505.24392}
  {arXiv:2505.24392 [quant-ph]} \BibitemShut {NoStop}%
\end{thebibliography}%

\end{document}